%% file: MuonFacility_snowmass22.tex
\documentclass[11pt, tightenlines, aps, prd, onecolumn, groupedaddress, superscriptaddress]{revtex4-2}
\pdfoutput=1
\usepackage{graphicx}
\usepackage{float}
\usepackage{amssymb}
\usepackage{amsmath}
\usepackage{array}
\usepackage{lineno}
\usepackage{scrextend}
\usepackage{siunitx}
\usepackage{textcomp}
\usepackage{gensymb}
\usepackage{hyperref}
\hypersetup{
    colorlinks=true,
    citecolor=blue,
    linkcolor=blue,
    filecolor=magenta,      
    urlcolor=blue,
    pdftitle={Sharelatex Example},
    }
\input{workshopsymbols}

\newcommand\snowmass{\begin{center}\rule[-0.2in]{\hsize}{0.01in}\\\rule{\hsize}{0.01in}\\
\vskip 0.1in Submitted to the Proceedings of the US Community Study\\ 
on the Future of Particle Physics (Snowmass 2021)\\ 
\rule{\hsize}{0.01in}\\\rule[+0.2in]{\hsize}{0.01in} \end{center}}

\begin{document}
\title{A New Charged Lepton Flavor Violation Program at Fermilab}

\input{author.tex}

\date{\today}
\begin{abstract}
The muon has played a central role in establishing the Standard Model of particle physics, and continues to provide valuable information about the nature of new physics. A new complex at Fermilab, the Advanced Muon Facility, would provide the world's most intense positive and negative muon beams by exploiting the full potential of PIP-II and the Booster upgrade. This facility would enable a broad muon physics program, including studies of charged lepton flavor violation, muonium-antimuonium transitions, a storage ring muon EDM experiment, and muon spin rotation experiments. This document describes a staged realization of this complex, together with a series of next-generation experiments to search for charged lepton flavor violation.
\snowmass
\end{abstract}

\maketitle

\def\thefootnote{\fnsymbol{footnote}}
\setcounter{footnote}{0}

\newpage 
\input{Sections/Summary}
\newpage
\input{Sections/Introduction}

\input{Sections/Theory}
\input{Sections/CompressorRing}

\input{Sections/Target}
\input{Sections/FFA}

\input{Sections/linac}

\input{Sections/MuDecay}

\input{Sections/MuConversion}

\input{Sections/Synergies}

\input{Sections/Conclusion}

\begin{acknowledgments}
This work was partly supported by the Fermi National Accelerator Laboratory, managed and operated by Fermi Research Alliance, LLC under Contract No. DE-AC02-07CH11359 with the U.S. Department of Energy. BE and DH are supported by the US Department of Energy under grant DE-SC0011925.  KL is supported by the US Department of Energy under grand DE-SC0019027.

\end{acknowledgments}

\bibliography{Bibliography.bib}

\end{document}

%% file: workshopsymbols.tex
\def\beq{\begin{equation}}
\def\eeq#1{\label{#1}\end{equation}}
\def\eeqn{\end{equation}}

\newenvironment{Eqnarray}%
   {\arraycolsep 0.14em\begin{eqnarray}}{\end{eqnarray}}
\def\beqa{\begin{Eqnarray}}
\def\eeqa#1{\label{#1}\end{Eqnarray}}
\def\eeqan{\end{Eqnarray}}

\def\lsim{\mathrel{\raise.3ex\hbox{$<$\kern-.75em\lower1ex\hbox{$\sim$}}}}
\def\gsim{\mathrel{\raise.3ex\hbox{$>$\kern-.75em\lower1ex\hbox{$\sim$}}}}

\def\del{\partial}
\def\Dslash{\not{\hbox{\kern-4pt $D$}}}
\def\dslash{\not{\hbox{\kern-2pt $\del$}}}
\def\pslash{\not{\hbox{\kern-2pt $p$}}}
\def\ETmiss{\not{\hbox{\kern-4pt $E$}}_T}

\def\Dlr{\mathrel{\raise1.5ex\hbox{$\leftrightarrow$\kern-1em\lower1.5ex\hbox{$D$}}}}

\def\MSB{{\bar{M \kern -2pt S}}}
\def\msb{{\bar{\scriptsize M \kern -1pt S}}}

\def\drb{{\bar{\scriptsize D \kern -1pt R}}}

\def\MeV{{\rm MeV}}
\def\GeV{{\rm GeV}}
\def\TeV{{\rm TeV}}



%% file: author.tex
\author{M.\ Aoki}
\affiliation{Graduate School of Science, Osaka University, 1-1 Machikaneyama, Toyonaka, Osaka, 560-0043, Japan}

\author{R. B.\ Appleby}
\affiliation{The University of Manchester, Department of Physics and Astronomy,Oxford Road,Manchester,M13 9PL, United Kingdom}
\affiliation{Cockcroft Institute, Sci-Tech Daresbury, Keckwick Lane, Daresbury, Warrington, WA4 4AD, United Kingdom}

\author{M.\ Aslaninejad}
\affiliation{School of Particles and Accelerators, Institute for Research in Fundamental Sciences (IPM), P.O.\  Box 19395-5531, Tehran, Iran}

\author{R.\ Barlow}
\affiliation{The University of Huddersfield, Queensgate, Huddersfield, HD1 3DH,UK}

\author{R.H.\ Bernstein}
\affiliation{Fermi National Accelerator Laboratory, P.O.\  Box 500, Batavia, IL 60510, USA}

\author{C.\ Bloise}
\affiliation{Laboratori Nazionali di Frascati dell’INFN, Via Enrico Fermi 40, 00044, Frascati, Italy}

\author{L.\ Calibbi}
\affiliation{School of Physics, Nankai University, Tianjin 300071, China}

\author{F.\ Cervelli}
\affiliation{INFN Sezione di Pisa
Ed.\ C Polo Fibonacci, Largo Pontecorvo 3, Pisa, Italy} 

\author{R.\ Culbertson}
\affiliation{Fermi National Accelerator Laboratory, P.O.\  Box 500, Batavia, IL 60510, USA}

\author{Andr\'{e} Luiz de Gouv\^{e}a}
\affiliation{Northwestern University,Department of Physics and Astronomy,
2145 Sheridan Road,
Evanston, IL 60208, USA}

\author{S.\ Di Falco}
\affiliation{INFN Sezione di Pisa
Ed.\ C Polo Fibonacci, Largo Pontecorvo 3, Pisa, Italy} 

\author{E.\ Diociaiuti}
\affiliation{Laboratori Nazionali di Frascati dell’INFN, Via Enrico Fermi 40, 00044, Frascati, Italy}

\author{S.\ Donati}
\affiliation{INFN Sezione di Pisa
Ed.\ C Polo Fibonacci, Largo Pontecorvo 3, Pisa, Italy} 

\author{R.\ Donghia}
\affiliation{Laboratori Nazionali di Frascati dell’INFN, Via Enrico Fermi 40, 00044, Frascati, Italy}

\author{B.\ Echenard}
\affiliation{California Institute of Technology, Pasadena, California 91125 USA}
\email[corresponding authors:]{echenard@caltech.edu}

\author{A.\ Gaponenko}
\affiliation{Fermi National Accelerator Laboratory, P.O.\  Box 500, Batavia, IL 60510, USA}

\author{S.\ Giovannella}
\affiliation{Laboratori Nazionali di Frascati dell’INFN, Via Enrico Fermi 40, 00044, Frascati, Italy}

\author{C.\ Group}
\affiliation{Department of Physics, University of Virginia, Charlottesville, Virginia 22904, USA}

\author{F. Happacher}
\affiliation{Laboratori Nazionali di Frascati dell’INFN, Via Enrico Fermi 40, 00044, Frascati, Italy}

\author{M.\ T.\ Hedges}
\affiliation{Department of Physics and Astronomy, Purdue University, 525 Northwestern Avenue, West Lafayette, IN 47907, USA}

\author{D.G.\ Hitlin}
\affiliation{California Institute of Technology, Pasadena, California 91125 USA}

\author{E.\ Hungerford}
\affiliation{Department of Physics, University of Houston, Houston TX,77204 USA}

\author{C.\ Johnstone}
\affiliation{Fermi National Accelerator Laboratory, P.O.\  Box 500, Batavia, IL 60510, USA}

\author{D.\ M.\ Kaplan}
\affiliation{Illinois Institute of Technology, 10 West 35th Street
Chicago, IL 60616, USA}

\author{M.\ Kargiantoulakis}
\affiliation{Fermi National Accelerator Laboratory, P.O.\  Box 500, Batavia, IL 60510, USA}

\author{D. J.\ Kelliher}
\affiliation{ISIS, STFC Rutherford Appleton Laboratory, Didcot OX11 0QX, UK}

\author{K.\ Kirch}
\affiliation{ETH Zurich Dep.\ Physik, Otto-Stern-Weg 1, 8093 Zurich, Switzerland}
\altaffiliation{Paul Scherrer Institute, Villigen, Switzerland
}

\author{A.\ Knecht}
\affiliation{Paul Scherrer Institute, Villigen, Switzerland
}

\author{Y.\ Kuno}
\affiliation{Graduate School of Science, Osaka University, 1-1 Machikaneyama, Toyonaka, Osaka, 560-0043, Japan}
\affiliation{Research Center of Nuclear Physics, Osaka University, 1-1 Yamadaoka, Suita, Osaka, 565-0871, Japan}

\author{A.\ Kurup}
\affiliation{Imperial College London, Exhibition Road, London SW7 2AZ, UK}
\author{J.-B.\ Lagrange}
\affiliation{ISIS, STFC Rutherford Appleton Laboratory, Didcot OX11 0QX, UK}

\author{M.\ Lancaster}
\affiliation{The University of Manchester,Department of Physics and Astronomy, Oxford Road, Manchester, M13 9PL, United Kingdom}

\author{K.\ Long}
\affiliation{Imperial College London, Exhibition Road, London SW7 2AZ, UK}

\author{A.\ Luca}
\affiliation{Fermi National Accelerator Laboratory, P.O.\  Box 500, Batavia, IL 60510, USA}

\author{K.\ Lynch}
\affiliation{York College and the Graduate Center, CUNY, New York, NY 11451, USA}

\author{S.\ Machida}
\affiliation{ISIS, STFC Rutherford Appleton Laboratory, Didcot OX11 0QX, UK}

\author{M.\ Martini}
\altaffiliation{Laboratori Nazionali di Frascati dell’INFN, Via Enrico Fermi 40, 00044, Frascati, Italy}
\affiliation{Universit\`{a} degli Studi Guglielmo Marconi, 00193, Rome, Italy}

\author{S.\ Middleton}
\affiliation{California Institute of Technology, Pasadena, California 91125 USA}

\author{S.\ Mihara}
\affiliation{Institute of Particle and Nuclear Studies (IPNS), KEK, Tsukuba, Ibaraki, 305-0801, Japan}

\author{J.\ Miller}
\affiliation{Boston University, 590 Commonwealth Ave., Boston MA 02215, USA}

\author{S.\ Miscetti}
\affiliation{Laboratori Nazionali di Frascati dell’INFN, Via Enrico Fermi 40, 00044, Frascati, Italy}

\author{L.\ Morescalchi}
\affiliation{INFN Sezione di Pisa
Ed.\ C Polo Fibonacci, Largo Pontecorvo 3, Pisa, Italy} 

\author{Y.\ Mori}
\affiliation{Institute for Integrated Radiation and Nuclear Science
Department of Nuclear Engineering, Kyoto University, Kyoto, Japan}

\author{P.\  Murat}
\affiliation{Fermi National Accelerator Laboratory, P.O.\  Box 500, Batavia, IL 60510, USA}

\author{B.\ Muratori}
\affiliation{ASTeC, STFC Daresbury Laboratory, Daresbury, Warrington, WA4 4AD Cheshire, United Kingdom}
\affiliation{Cockcroft Institute, Sci-Tech Daresbury, Keckwick Lane, Daresbury, Warrington, WA4 4AD, United Kingdom}

\author{D.\ Neuffer}
\affiliation{Fermi National Accelerator Laboratory, P.O.\  Box 500, Batavia, IL 60510, USA}

\author{A.\ Papa}
\affiliation{INFN Sezione di Pisa
Ed.\ C Polo Fibonacci, Largo Pontecorvo 3, Pisa, Italy} 

\author{J.\ Pasternak}
\affiliation{Imperial College London, Exhibition Road, London SW7 2AZ, UK}
\altaffiliation{ISIS, STFC Rutherford Appleton Laboratory, Didcot OX11, 0QX, UK}

\author{E.\ Pedreschi}
\affiliation{INFN Sezione di Pisa
Ed.\ C Polo Fibonacci, Largo Pontecorvo 3, Pisa, Italy} 

\author{G.\ Pezzullo}
\affiliation{Department of Physics, Yale University, 56 Hillhouse, New Haven, CT-06511, USA}

\author{T.\ Planche}
\affiliation{TRIUMF, 4004 Wesbrook Mall, Vancouver, BC, Canada V6T 2A3}

\author{F.\ Porter}
\affiliation{California Institute of Technology, Pasadena, California 91125 USA}

\author{E.\ Prebys}
\affiliation{UC Davis, Department of Physics and Astronomy, 
One Shields Avenue 
Davis, CA 95616 }

\author{C. R.\ Prior}
\affiliation{ISIS, STFC Rutherford Appleton Laboratory, Didcot OX11 0QX, UK}

\author{V.\ Pronskikh}
\affiliation{Fermi National Accelerator Laboratory, P.O.\  Box 500, Batavia, IL 60510, USA}

\author{R.\ Ray}
\affiliation{Fermi National Accelerator Laboratory, P.O.\  Box 500, Batavia, IL 60510, USA}

\author{F.\ Renga}
\affiliation{Istituto Nazionale di Fisica Nucleare, Sez.\ di Roma, P.\ le A. Moro 2, 00185 Roma, Italy}

\author{C.\ Rogers}
\affiliation{ISIS, STFC Rutherford Appleton Laboratory, Didcot OX11 0QX, UK}

\author{I.\ Sarra}
\affiliation{Laboratori Nazionali di Frascati dell’INFN, Via Enrico Fermi 40, 00044, Frascati, Italy}

\author{A.\ Sato}
\affiliation{Graduate School of Science, Osaka University, 1-1 Machikaneyama, Toyonaka, Osaka, 560-0043, Japan}

\author{S. L.\ Smith}
\affiliation{ASTeC, STFC Daresbury Laboratory, Daresbury, Warrington, WA4 4AD Cheshire, United Kingdom}
\affiliation{Cockcroft Institute, Sci-Tech Daresbury, Keckwick Lane, Daresbury, Warrington, WA4 4AD, United Kingdom}

\author{F.\ Spinella}
\affiliation{INFN Sezione di Pisa
Ed.\ C Polo Fibonacci, Largo Pontecorvo 3, Pisa, Italy} 

\author{D.\ Stratakis}
\affiliation{Fermi National Accelerator Laboratory, P.O.\  Box 500, Batavia, IL 60510, USA}

\author{M.\ Syphers}
\affiliation{Northern Illinois University, DeKalb, IL 60115, USA}
\altaffiliation{Fermi National Accelerator Laboratory, Batavia IL 60510, USA}

\author{N.M.\ Truong}
\affiliation{UC Davis, Department of Physics and Astronomy, 
One Shields Avenue 
Davis, CA 95616 }

\author{S.\ Tygier}
\affiliation{The University of Manchester, Department of Physics and Astronomy,Oxford Road,Manchester,M13 9PL, United Kingdom}
\affiliation{Cockcroft Institute, Sci-Tech Daresbury, Keckwick Lane, Daresbury, Warrington, WA4 4AD, United Kingdom}

\author{Y.\ Uchida}
\affiliation{Imperial College London, Exhibition Road, London SW7 2AZ, UK}

\author{M.\  Yucel}
\affiliation{Fermi National Accelerator Laboratory, P.O.\  Box 500, Batavia, IL 60510, USA}

%% file: Sections/Summary.tex
\section{Executive summary}
\label{sec:execsummary}

This report outlines an opportunity to create a next-generation muon facility at Fermilab to deliver the world’s most intense positive and negative muon beams. The Advanced Muon Facility (AMF) would exploit the full potential of the PIP-II accelerator to explore muon physics with unprecedented sensitivity: charged lepton flavor violation, muonium-antimuonium transitions, a storage ring muon EDM experiment or muon spin rotation are just some of the possibilities. This project has also strong synergies with the development of a muon collider and a new Dark Matter program at Fermilab (with its own strong physics case). This document proposes a staged realization of this complex, and discuss the physics potential to further explore charged lepton flavor violation (CLFV). 

The current generation of CLFV studies will search for new physics (NP) in rare muon processes at mass scales $\geq 10^4~\TeV$. These experiments, underway or in preparation at Fermilab, PSI, and J-PARC, will have sensitivity up to four orders of magnitude better than current limits. This is made possible by intense low energy muon beams and experiments designed to function at the resulting high rates. 

The sensitivity of these experiments will take us to a regime in which many New Physics models predict observable CLFV signals. Observation of these charged lepton flavor violating processes would be unambiguous evidence of physics beyond the Standard Model. If they are not seen, higher sensitivity searches will have to be devised. If CLFV processes are found, detailed studies of the rates of differing processes along with the atomic number dependence of muon-to-electron conversion will be extremely important, requiring a further increase in the intensity and quality of muon beams and their accompanying experiments. AMF will provide intensities two orders of magnitude beyond those currently available, providing sensitivities of the order of $10^{-19}$ or better for muon-to-electron conversion with a proton beam power of ${\cal O}(1)$ MW. 

The Advanced Muon Facility takes advantage of the fact that the neutrino program requires only a small fraction of the total power available at PIP-II. Up to 1.6 MW is available to produce the required intense muon beams. AMF consists of a 50\thinspace m compressor ring to provide a suitable time structure for the experiments, a high power target, a superconducting production solenoid to gather pions and muons, and a small fixed field alternating gradient accelerator (FFA) that produces a low momentum beam of muons with very well-defined momentum (about 1\%). The low momentum is important to ensure that the bulk of the muons can be brought to rest in an extremely thin target. Design studies are underway to understand how low a muon momentum (around $30~\MeV$ is ideal) can be built. If the result is higher, a small induction linac can be added to further reduce the central muon energy. The facility could deliver both negative and positive muon beams, enabling a full suite of muon decay and conversion experiments, and we are investigating providing both signs simultaneously.

This facility would provide the foundation for a comprehensive muon physics program in the next decade and beyond. The PIP-II complex is already underway, but only a small fraction of its power is used by the future neutrino program. This proposal represents a unique opportunity to fully exploit the remaining capabilities, complementing the neutrino program with an on-site world-class physics program.

%% file: Sections/Introduction.tex
\section{Introduction}
\label{sec:introduction}

The muon has always played a central role in understanding the structure of the weak interaction and establishing the Standard Model (SM) of particle physics -- its discovery was the first evidence for flavor and generations. Building on this legacy, muon physics continues to provide valuable information about the nature of the dynamics beyond the Standard model (SM). One exceptionally promising avenue is the search for rare processes violating charged lepton flavor  which, if observed, would be an unambiguous sign of New Physics (NP). Measurements of precisely known quantities in the SM, such as the anomalous magnetic moment of the muon, provide another avenue to probe NP. In both cases, the availability of intense muon sources is critical to achieve the greatest potential for discoveries.

This document calls attention to an opportunity to create a next-generation muon facility at Fermilab that would deliver the world's most intense $\mu^+$ and $\mu^-$ beams. The Advanced Muon Facility (AMF) would exploit the full potential of the PIP-II accelerator to explore muon physics with unprecedented sensitivity: charged lepton flavor violation~\cite{CLFV:LOI}, muonium-antimuonium transitions~\cite{conlin2020,willmann1999}, a storage ring muon EDM experiment~\cite{Semertzidis:2016wtd}, or $\mu$ spin rotation~\cite{Kaplan:LOI} are just some of the possibilities. Furthermore, this program has many synergistic activities with R\&D efforts to develop a muon collider or a future dark matter program at FNAL. 

 We will focus the discussion on charged lepton flavor violating (CLFV) muon processes; other physics topics will be covered in subsequent publications. Among all CLFV reactions, those involving muons generally offer the best sensitivity at low energies, and several experimental efforts are underway to study them. MEG-II~\cite{MEGII2018} and Mu3e~\cite{BERGER201435} at PSI are expected to probe the $\mu^+ \rightarrow e^+ \gamma$ and $\mu^+ \rightarrow e^+e^-e^+$ decay channels at the level of $6\times 10^{-14}$ and $\sim 10^{-15}$, respectively. The proposed High Intensity Muon Beam (HiMB) at PSI~\cite{Aiba:2021bxe} could further increase these sensitivities by an order of magnitude. On the conversion side, both Mu2e at FNAL~\cite{bartoszek2015mu2e} and COMET at J-PARC~\cite{comet-tdr} are expected to probe conversion rates in the vicinity of $10^{-17}$ for Al targets. 

Dedicated experiments at AMF could potentially improve the sensitivity of decay channels by two orders of magnitude compared to the rates probed with the HiMB, and reach conversion rates down to the level of $10^{-19}$ or better with a proton beam power of ${\cal O}(1)$ MW. This level of sensitivity will explore effective mass scales up to $\sim 10^{4}$\thinspace TeV/c$^2$, well above the direct detection limit of colliders. This complex would also enable the study of muon conversion with high-$Z$ target materials, which could provide critical information about the nature of the underlying New Physics~\cite{Cirigliano:2009bz}. The statistics are so high that the experiments could even consider using low-$Z$ materials such as Li, opening new searches not easily available to heavier targets.

\begin{figure}[h]
 \begin{center}
\includegraphics[width=0.7\textwidth]{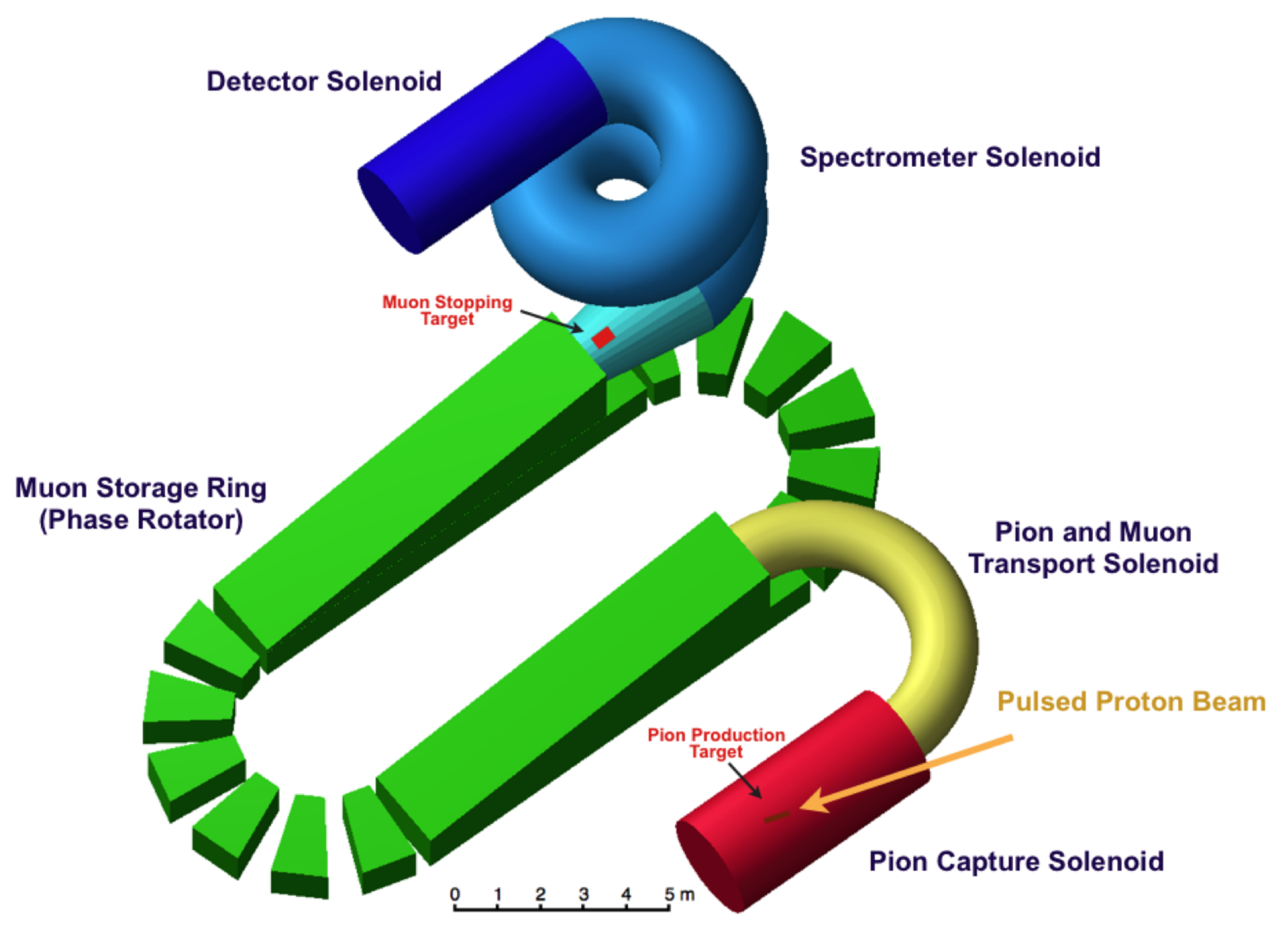}
\end{center}
    \caption{The PRISM concept, adapted from Ref.~\cite{KUNO2005376}, showing the facility configured for muon conversion experiments.  Not shown are the PIP-II linac, the RF beam splitter and transport lines, the compressor ring, and the induction linac.  The spectrometer and detector solenoids could be replaced for upgrades or new, different experiments.}
    \label{fig:prism}
\end{figure}

The facility itself is based on  a small fixed-field alternating gradient synchrotron (FFA), used to produce a cold, intense muon beam with well-defined momentum. The PRISM (Phase Rotated Intense Source of Muons) system~\cite{KUNO2005376}, shown in Fig.~\ref{fig:prism}, provides a reference concept. Short high intensity proton bunches are delivered to a production target surrounded by a capture solenoid with a field at about 5T, well within current capabilities. The muons produced by pion decays are then injected into the FFA ring by a transport system. The phase rotation decreases the momentum spread of the muons, trading time spread for momentum spread. During the RF phase rotation, the remaining pion contamination is reduced to negligible levels (since the  pions decay quickly.)  A cold quasi-monochromatic muon beam is then extracted to the detector system. The feasibility of the FFA approach was demonstrated with a dedicated prototype at the Research Center of Nuclear Physics (RCNP) of Osaka University~\cite{Witte:2012zza}. The muon beam leaving the FFA can be further reduced in momentum, if required, with a small induction linac.

The phase rotation requires a very short proton pulse, ${\cal O}(10\rm\, ns)$, and a compressor ring is needed to rebunch the protons from PIP-II to create the required beam structure. A major limiting factor of this scheme is the beam power that can be absorbed by the target. Several concepts have been developed to handle a beam power of ${\cal O}(100)$ kW in a capture solenoid, but additional R\&D is needed to design a MW-class target. A similar challenge is faced by the muon collider~\cite{calvianiTalk}; the two efforts are potentially synergistic, with the target required for this program representing both a staging and an R\&D platform for the demands of the muon collider.

We present a phased implementation to realize the full potential of AMF: a conceptual design for a complex with a $\sim 100$ kW proton beam power using PIP-II as a first stage, followed by an upgrade to reach a final $\sim 1$ MW power. A 100 kW facility would already provide significant sensitivity improvements for both decay and conversion experiments, and allow measurement of conversion in high-$Z$ materials, currently inaccessible with pulsed-beam experiments. We also outline the challenges to the design of these experiments, and the potential approaches to overcome these challenges. If we start design studies in the near future, this program could be realized after the completion of the Mu2e experiment on the FNAL site, and operate simultaneously with LBNF/DUNE.

This paper is organized as follows. The physics of muon CLFV is briefly reviewed in  Section~\ref{sec:physics}, followed by an discussion of PIP-II, the compressor ring, and the target system in Sections~\ref{sec:compressor} and \ref{sec:target}. The FFA is presented in Section~\ref{sec:FFA}, and the possibility of adding an induction linac to further slow the muon beam is discussed in Section~\ref{sec:linac}. Considerations on decay and conversion experiments are presented in Sections~\ref{sec:mudecay} and \ref{sec:conversion}, and synergies with other experimental efforts are examined in Section~\ref{sec:synergies}. Potential design improvements, such as the possibility to produce $\mu^+$ and $\mu^-$ beams simultaneously, are also outlined throughout the document.

%% file: Sections/Theory.tex
\section{Physics of Charged Lepton Flavor Violation}
\label{sec:physics}

Lepton flavor is a conserved quantity in the Standard Model with massless neutrinos. The introduction of non-zero neutrino masses and mixing angles, following the observation of neutrino oscillations, provides a mechanism for charged lepton flavor violation (CLFV) in the SM through mixing in loops. If neutrino masses arises from Yukawa interactions with the Higgs boson, the CLFV rates are typically GIM-suppressed by factors of $\sum_{i,j}(\Delta m_{i,j}^2 / M_W^2)^2$, where $\Delta m^2$ is the mass-squared difference between  the $i$-th and $j$-th neutrino mass eigenstates. For $\mu \rightarrow e \gamma$, the corresponding branching fraction is of the order of $10^{-54}$~\cite{Petcov:1976ff}, well below the experimental sensitivity of any practical  experiment. Many scenarios of physics beyond the Standard Model introduce new sources of CLFV, leading to rates that are potentially accessible to next generation experiments (see e.g. Ref.~\cite{Calibbi:2017uvl, Cei:2014jtm}). An observation of CLFV is therefore an unambiguous sign of New Physics. Furthermore, CLFV is closely linked to the physics of neutrino masses and the prevalence of matter in the universe through leptogenesis. Several explanations of the observed neutrino masses involve heavy neutrino partners (see e.g. Ref~\cite{Hambye:2013jsa}), and predict large CLFV effects. Together with neutrino measurements, CLFV reactions can strongly constrain these models and open a portal into GUT-scale physics. 

While a large variety of CLFV processes can be studied, the muon sector has attracted much attention due to the availability of intense sources and the relatively long muon lifetime; the intense sources provide statistics, and the long muon lifetime makes it possible to produce clean beams for experiments. Three main transitions have been investigated so far: $\mu^+ \rightarrow e^+ \gamma$, $\mu^+ \rightarrow e^+e^-e^+$, and $\mu^-N \rightarrow e^-N$ conversion in the Coulomb field of a nucleus. In addition to  excellent experimental sensitivity, these channels provide complementary information about the nature of new CLFV sources. In scenarios dominated by the contribution of penguin diagrams, such as many supersymmetric setups~\cite{Calibbi:2017uvl} or the scotogenic model~\cite{Ma:2006km}, $\mu^+ \rightarrow e^+ \gamma$ rates can be several orders of magnitude larger than those of $\mu^+ \rightarrow e^+e^-e^+$ decays and $\mu^- - e^-$ conversion. Alternatively, tree level interactions, including leptoquarks~\cite{Baek:2015mea, Hati:2020cyn} and $\mathrm{Z}'$ models~\cite{Crivellin:2015era}, are best probed with $\mu^+ \rightarrow e^+e^-e^+$ decays and $\mu^- - e^-$ conversion as $\mu^+ \rightarrow e^+ \gamma$ rates are (strongly) suppressed. 

\begin{figure}
    \centering
    \includegraphics[width=0.5\textwidth]{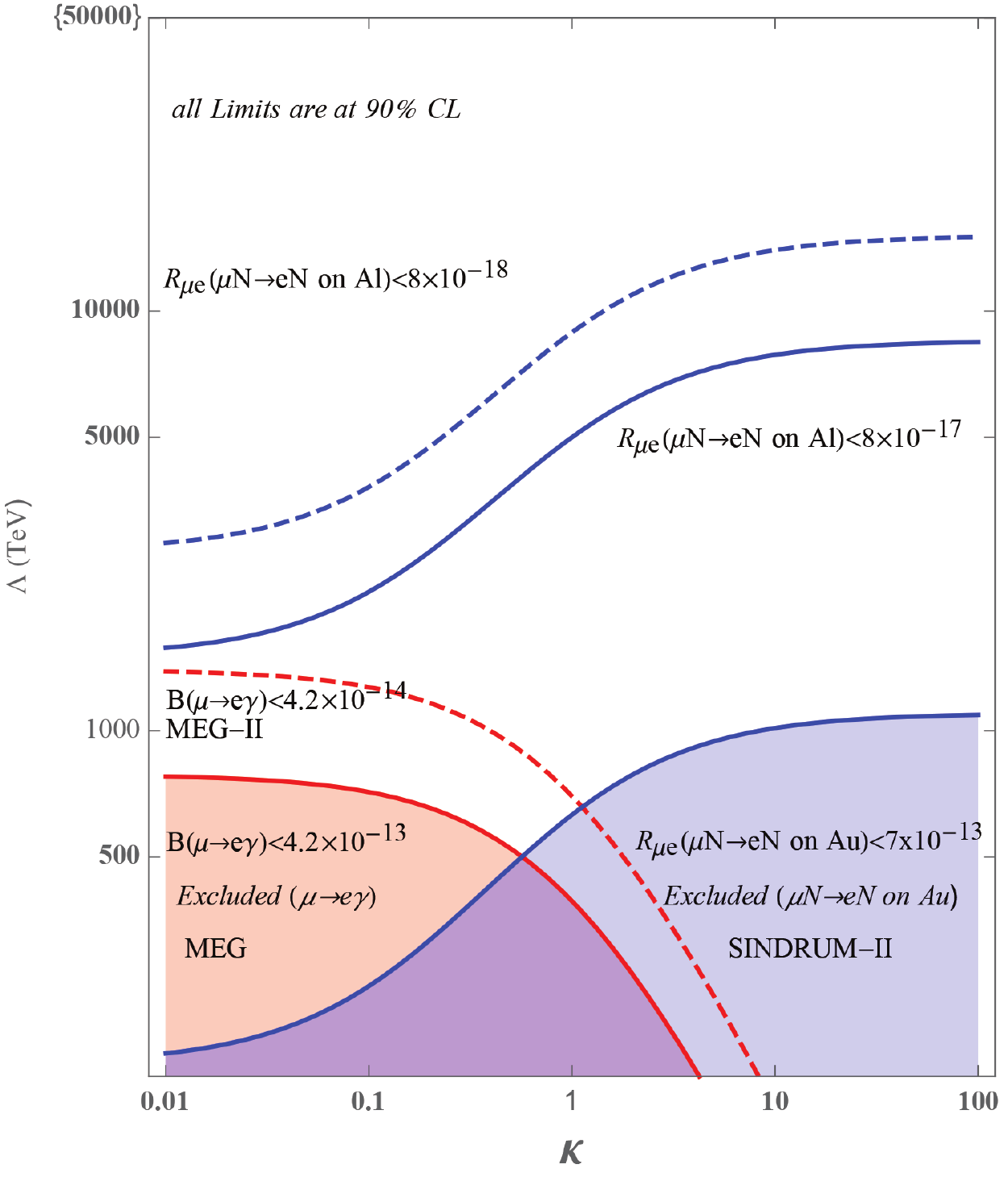}
    \caption{Sensitivity of a $\mu^- N \rightarrow e^- N$ conversion in $^{27}$Al and to a $\mu \rightarrow e \gamma$ search to the New Physics scale $\Lambda$ as a function of $\kappa$, as defined in Eqn.~\ref{eqn:deGouveaLagrangian} together with existing experimental bounds. Adapted from Ref.~\cite{deGouvea}.}
    \label{fig:deGouveaPlot}
\end{figure}

Adopting a more generic point of view, a simple effective Lagrangian has been proposed to illustrate the sensitivity to new CLFV effects in $\mu \rightarrow e \gamma$ and $\mu-e$ conversion~\cite{deGouvea}:
\begin{equation}
{\cal L}_{ {\rm CLFV} } = \frac{m_{\mu}}{(\kappa + 1) \Lambda^2} \bar{\mu}_R \sigma_{\mu\nu} e_L F^{\mu \nu} + \frac{\kappa}{(\kappa + 1) \Lambda^2}
\bar{\mu}_L \gamma_{\mu} e_L (\bar{u}_L \gamma^{\mu} u_L + \bar{d}_L \gamma^{\mu} d_L ) \label{eqn:deGouveaLagrangian}
\end{equation}
where $\kappa$ denotes the relative strength between the two terms. The $\mu \rightarrow e \gamma$ decay typically probes the region $\kappa \ll 1$, for which dipole-type operators dominate, while $\mu-e$ conversion is also sensitive to four-fermion operators ($\kappa \gg 1$); the mass scale reached as a function of $\kappa$ for this toy Lagrangian  is illustrated in Fig.~\ref{fig:deGouveaPlot}. The current experimental limits probe NP scales at the level of $\Lambda \sim 10^3 \, \TeV$ for the relevant operators. A similar effective Lagrangian with purely leptonic terms can also be constructed to analyze the sensitivity of $\mu \rightarrow eee$ decays. Similarly to muon conversion, this transition receives contributions from both type of operators, but only probe mass scales of the order of $\sim 10^{2} \, \TeV$ with the existing experimental bounds.    

 Currently planned experiments will significantly increase these sensitivities. MEG-II plans to improve the $\mu \rightarrow e \gamma$ bound by an order of magnitude, while the Mu3e experiments should ultimately reach $\mu \rightarrow eee$ branching fractions at the level of $10^{-16}$ with the High Intensity Muon Beam upgrade at PSI. Both Mu2e at FNAL and COMET at J-PARC plan to increase the reach of $\mu-e$ conversion by four orders of magnitude. These outstanding sensitivities will explore NP mass scales at the level of $10^3 - 10^4 \, \TeV$, well beyond the reach directly accessible at colliders.

\begin{figure}
  \centering
  \includegraphics[width=0.55\textwidth]{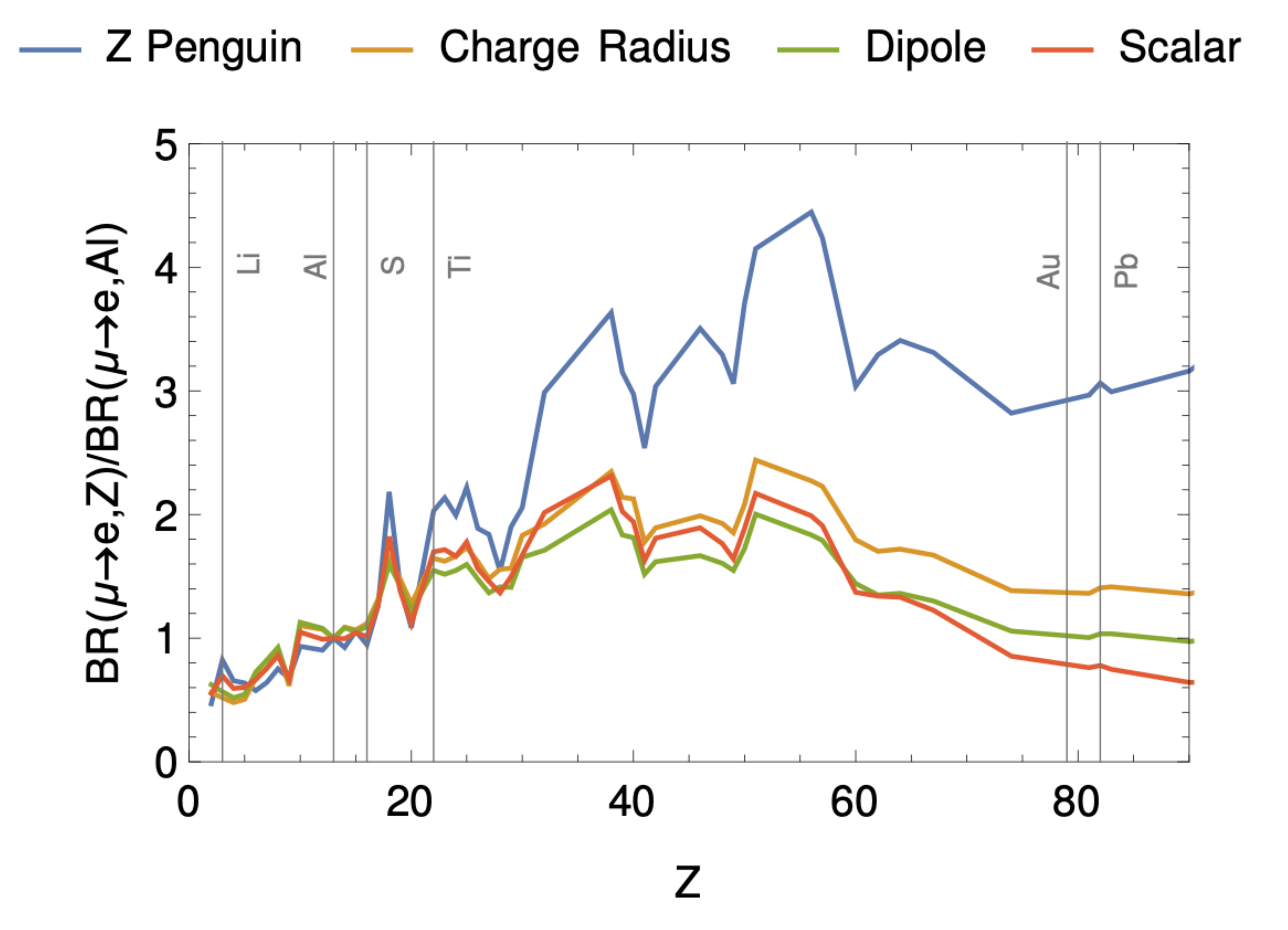}
  \includegraphics[width=0.4\textwidth]{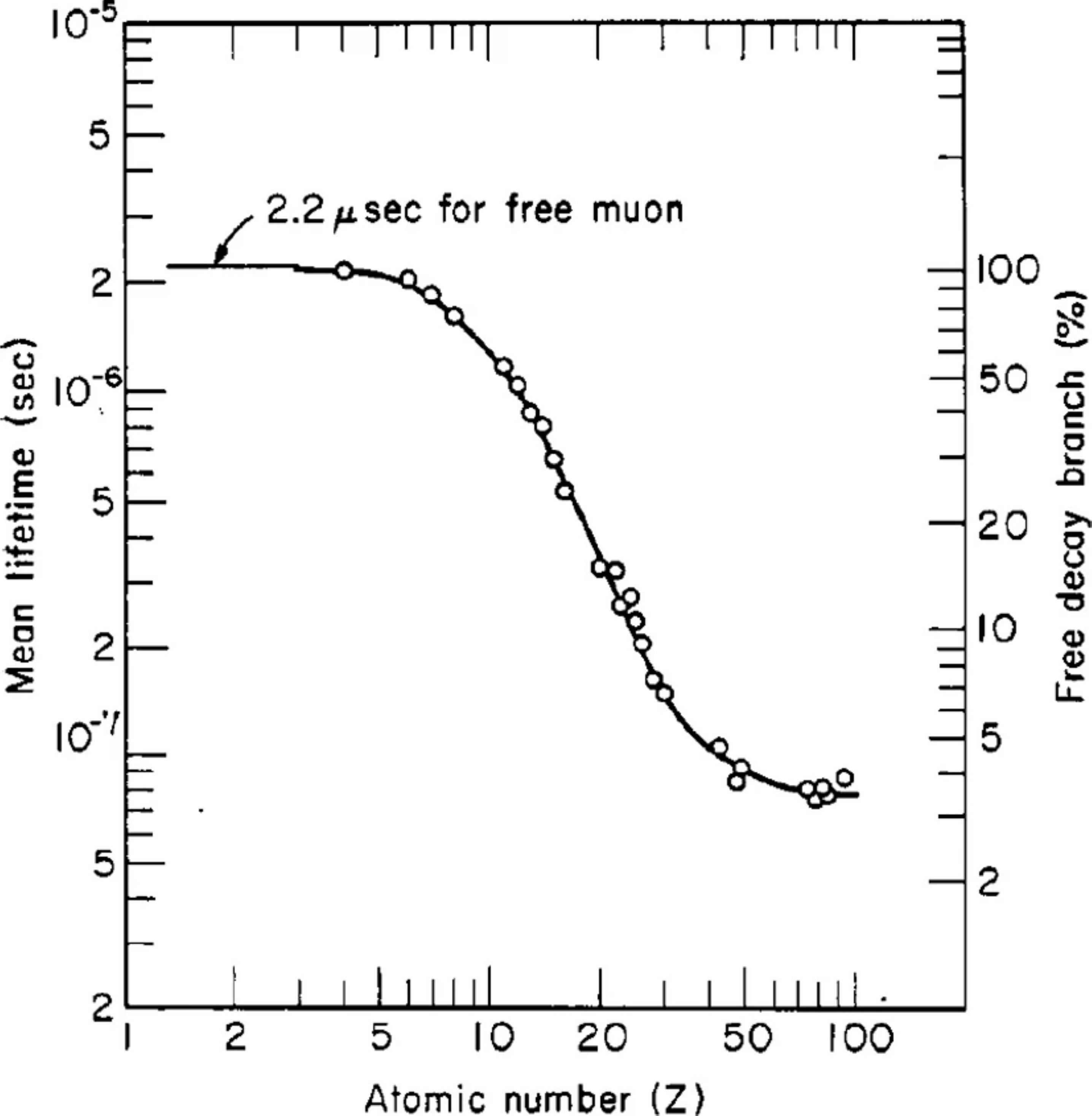}
\caption{(left) $Z$ dependence of $\mu$-conversion rates for different types of operators modeling New Physics effects, and (right) mean muonic atom lifetime as a function of the atomic number. Adapted from Refs. \protect{\cite{Heeck:LOI, Knecht:2020}}, respectively.}
\label{fig:sens-and-lifetime-vs-z}
\end{figure}

The conversion process can also provide information about the underlying structure of New Physics by using different target materials~\cite{Cirigliano:2009bz}. As shown in Fig.~\ref{fig:sens-and-lifetime-vs-z}, the $Z$-dependence of the conversion rate can be used to discriminate between operators (dipole, scalar or vector) providing the dominant source of CLFV. The difference is generally larger for heavier nuclei, but the corresponding muonic atom lifetime is much shorter (Fig.~\ref{fig:sens-and-lifetime-vs-z}), imposing significant constraints on the experiment. Two light nuclei would still provide discriminating power, but the rates need to be determined with greater precision. 

A similar process violating both lepton number and lepton flavor, $\mu^- N(Z,A) \rightarrow e^+ N(Z-2,A)$, can only proceed if neutrinos have a Majorana mass term. This channel is complementary to neutrinoless double beta decay, probing transitions between different flavors instead of flavor-diagonal couplings. Upcoming experiments should significantly improve the bounds on $\mu^- - e^+$ conversion and probe the effective Majorana neutrino mass scale $\langle m_{e\mu} \rangle$ down to the MeV region~\cite{Yeo:2017fej}. 

Muon decays are also probe new light, weakly coupled particles, such as dark sectors with states below the muon mass. These states might have LFV coupling, such as axion-like particles, or could conserve flavor in the case of dark photons. They lead to a wide variety of signatures, including $\mu \rightarrow e X_{NP}$, $\mu \rightarrow e \gamma X_{NP}$ or $\mu \rightarrow e \nu\nu X_{NP}$, that could be investigated by the next generation of CLFV experiments~\cite{CLFVLightNP:LOI, Echenard:2014lma,  Calibbi:2020jvd}. 

In summary, CLFV experiments play a central role in the search for BSM physics by exploring uniquely accessible processes complementary to collider and neutrino experiments. If a signal is observed by MEG-II, Mu3e, Mu2e or COMET, studying the Z-dependence of the conversion rate will provide critical information about the structure of New Physics. If not, higher intensity muon beams will be required to further improve the search, probing higher mass scales and constraining models. The physics case for a next generation of experiments is well motivated in both cases. 

%% file: Sections/CompressorRing.tex
\section{PIP-II and compressor ring}
\label{sec:compressor}

PIP-II is an upgrade to the Fermilab accelerator complex to increase the beam power available for the high energy neutrino program~\cite{osti_1365571}.  The centerpiece of this project is a new, 800 MeV superconducting linac that will inject protons into the existing 8 GeV Booster, replacing the current 400 MeV linac.  The increased injection energy, coupled with a sophisticated transverse painting scheme, will allow the Booster to accelerate more beam (``painting" refers to spreading the beam out both transversely and longitudinally.)
In addition, the Booster repetition rate will be increased from 15 to 20 Hz.
Although the high energy neutrino program is the primary motivation for PIP-II, that program will use less than 1\% of its capacity. There will potentially be 1.6 MW of $800~\MeV$ beam available for other research opportunities.

Our initial plan, described elsewhere~\cite{mu2eii:WP}, is to use the fully configurable bunch structure of the PIP-II beam to mimic the 600 kHz bunch structure that the Mu2e experiment will get from the Delivery Ring. However, there is no way to directly use the beam from the PIP-II to load a FFA~\cite{Appleby2020}. To operate properly, the FFA will need  proton bunches that are on the order of 10 ns long at 100-1000 Hz (see Section~\ref{sec:FFA}), but the direct output of the PIP-II Linac would be minuscule with that structure.

Driving a FFA will require a ``compressor ring" to accumulate protons and extract them with the requisite bunch structure.   There are numerous challenges to building such a ring, with the most significant being space charge tune shift~\cite{Prebys2018}, which will limit the total number of protons that can be loaded.  For Gaussian hadron beams in a  circular synchrotron, the maximum space charge tune shift $\Delta\nu$ is given by
\begin{align*}
    \left|\Delta \nu\right| &= \frac{Bn_bN_br_0}{4\pi\beta\gamma^2\epsilon_N}
\end{align*}
where $n_b$ and $N_b$ are the number of bunches and the number in each bunch, respectively, $r_0$ is the classical proton radius, $\beta$ and $\gamma$ are the usual Lorentz factors, $\epsilon_N$ is the normalized emittance, and $B$ is the ``bunching factor" defined as the peak linear density over the average linear density. Normally, the tune shift is limited to about 0.2.

For a fixed energy, the maximum bunch size scales therefore as
\begin{align*}
    N_{b,max}&\propto \frac {\epsilon_N}{Bn_b}\\
    &= \frac{t_b}{\tau}\epsilon_N
\end{align*}
where $t_b$ and $\tau$ are the bunch length and period, respectively, in units of time, assuming a uniform linear distribution.

\begin{figure} [ht]
    \centering
    \includegraphics[width=5 in]{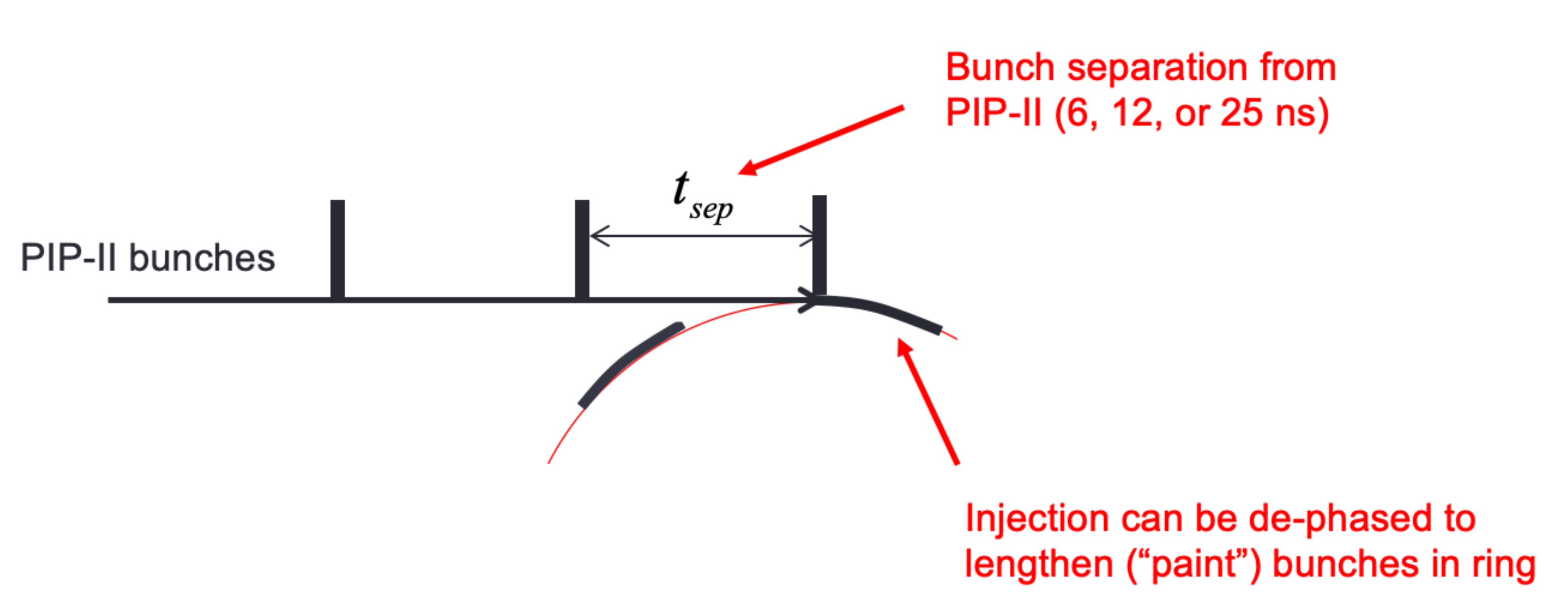}
    \caption{Schematic illustration of injection into the compressor ring. Beam is distributed both transversely and longitudinally.}
    \label{fig:compressor-injection}
\end{figure}

\begin{figure} [ht]
    \centering
    \includegraphics[width=6 in]{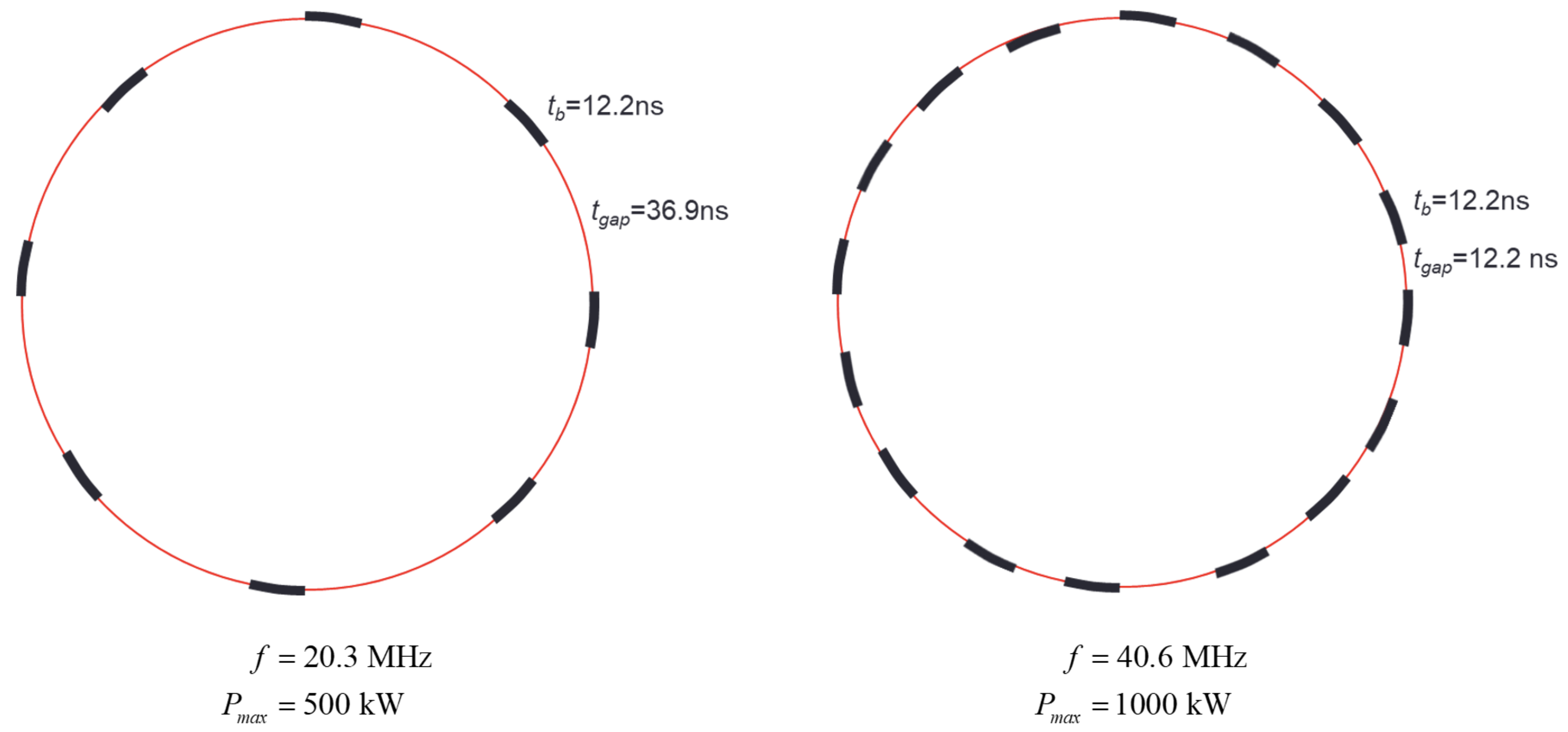}
    \caption{Two possibilities for a 100 m compressor ring. The ring on the left shows 20 MHz bunches, each 12 ns long, separated by 37 ns. This ring would reach a total power of $\sim500 \rm \, kW$. To achieve 1 MW, a bunch frequency of 41 MHz would be needed (right ring), but this would represent a challenge for the extraction kicker.}
    \label{fig:compressor-ring-options}
\end{figure}

The space charge tune shift can be mitigated by keeping the circumference of the ring as small as possible, as well as by painting the beam as illustrated in Fig.~\ref{fig:compressor-injection}. However, the desire for a small circumference puts the needs of this compressor ring at odds with a Booster ``pre-loader" compressor ring, which would have to be the same circumference as the Booster. 

Two options for a 100 m circumference compressor ring are shown in Fig.~\ref{fig:compressor-ring-options}. The left image shows bunches being loaded every eight PIP-II bunches (about 20 MHz). Loading is phased to stretch the bunches to about 12 ns, leaving about 37 ns between them. Loading would be staggered such that full bunches could be extracted at a constant rate.  Table~\ref{tab:compressor} shows the bunch size required for various beam powers, assuming a 100 Hz extraction rate, as well as the normalized emittance required to keep the space charge tune shift below about 0.2 for both a 100 m ring and the 500 m ring required for the Booster Storage Ring. In the case of the larger ring, the 500 kW and 1000 kW options would lead to beam sizes larger than the SNS, which would present a challenge for magnet design.

Unfortunately, the $2\times 10^8$ maximum bunch size out of the PIP-II linac limits the maximum power to 500 kW at 20 MHz, and a frequency of 41 MHz (every fourth PIP-II bunch) would be required to reach 1 MW, an option as shown at the right of Fig.~\ref{fig:compressor-ring-options}. This would only leave about 12 ns for the extraction kicker rise and fall, which could be challenging.

\begin{table}
    \centering
    \caption{Required normalized emittance to maintain a maximum tune shift ${}<0.2$ for rings of circumferences of 100 and 500 m at 100, 500, and 1000 kW total beam power.}
    \label{tab:compressor}
 \begin{tabular}{ l | c c  c  c  c  c }
& \multicolumn{3}{ c }{\bf C=100 m} & \multicolumn{3}{ c }{\bf C=500 m (BSR)} \\
{\bf Power [kW]} &  100 & 500 & 1000 & 100 & 500 & 1000 \\
\hline\hline
{ $N_b$ [$10^{12}$]} & 7.8 & 39.1 & 78.1 & 7.8 & 39.1 & 78.1 \\
{ $\epsilon_N$ [$\pi$-mm-mr] }& 54 & 268 & 536 & 268 & 1339 & 2678 \\
radius ($\beta_\perp$ = 20m) [mm] & 26	&58	&83	&58	&131	&185\\
 \end{tabular}
 \end{table}

%% file: Sections/Target.tex
\section{Production target}
\label{sec:target}

The initial conceptual muon production design for this facility is based on the target-in-solenoid proposal pioneered for the MELC experiment~\cite{osti_286496} and refined into the Mu2e~\cite{bartoszek2015mu2e} and COMET~\cite{comet-tdr} designs.  A production target is supported along the magnetic axis of a high-field solenoid, while the primary proton beam is brought in off-axis to intersect the target and produce the secondary beam.  The spent beam passes out the far end of the solenoid and is then collected in a beam dump.  In both the Mu2e and COMET designs, the muons either produced toward, or reflected by the graded field into, the backward direction are captured into a curved solenoidal transport system --- S-shaped in the case of Mu2e, and C-shaped in the case of COMET and this proposal. These curved solenoid designs naturally charge-separate the collected particles: positive species will drift normal to the curve plane in the opposite direction to negative species. At the end of a C-shaped transport channel, wrong-sign particles could be collimated away, or the charge-separated beams could be delivered independently for different experiments.

While our initial concept retains the backward production feature, the FFA will virtually eliminate the prompt backgrounds for CLFV experiments.  Thus, we are exploring alternate collection strategies that could enhance the muon rates at any beam power, or simplify target monitoring, maintenance, and replacement. Alternatives obviously include forward collection, but also transverse collection --- as in the PSI HiMB design --- and potentially more exotic possibilities such as hybrid toroid-solenoid options~\cite{DiBitonto:1974si}. The facility could even provide muon or pion beams simultaneously to different experiments from both ends of a conventional solenoid design. 

As mentioned in Section~\ref{sec:introduction}, we envision a phased development of this facility, first to $\sim 100~\rm kW$ of proton beam power, growing later to $\sim 1~\rm MW$, operating at 100-1000 Hz pulse rate. 
A R\&D effort for a 100 kW scale target operating with $800~\MeV$ beam from PIP-II is already underway for the parallel Mu2e-II proposal~\cite{mu2eii:WP}.  In particular, a Fermilab-funded LDRD collaboration~\cite{ref:Fang2020} is working on conceptual designs that may meet the needs for our low-power beam scenario.  The driving constraints are peak stresses, radiation damage, and heat management.  The core of the leading design addresses all of these challenges at the cost of mechanical complexity: balls of target material will be circulated through a piping system 
and intersect the beam path once per circulation cycle. The spherical balls will minimize damaging stresses. This conveyor target design will bring the balls outside the solenoid into a region with sufficiently low radiological exposure so that they can be replaced during operations without bringing the entire facility offline, as it would be the case with a monolithic target.  There are still a number of challenges to be addressed, for example efficient heat removal may require a phase-changing fluid. 

Scaling the target system to ${\cal O}$ (1 MW) will require a concerted R\&D effort.  Megawatt-scale targets are in operation or under construction around the world in neutrino production facilities and at spallation sources (some examples  in the U.S.\ include NOvA, LBNF, and SNS~\cite{NOvA:2007rmc, Papadimitriou:2018akk, Winder:2021baz}), so there is significant experience to draw on.  The key challenge for our facility comes from trying to embed a compact production target within the relatively small open bore of a superconducting solenoid  while protecting the solenoid from the heat and radiation load.  Muon collider design and prototyping efforts have attempted to address this problem with various systems --- liquid metal or granular jets, for example~\cite{calvianiTalk} --- but at larger size than we require, and with anticipation of routine replacement of the target and the solenoid shielding.  Many of these target material choices are now disfavored for environmental reasons, but current directions in muon collider R\&D provide a number of good starting points for our needs, demonstrating the synergies between the two efforts.

%% file: Sections/FFA.tex
\section{Fixed Field Alternating Gradient Synchrotron (FFA) }
\label{sec:FFA}

The scheme presented in this document is based on the PRISM system, which was proposed to produce the beam for a next generation muon-to-electron conversion experiment~\cite{KUNO2005376} using a Fixed Field Alternating gradient (FFA)~\cite{Symon:1956pr} ring, as shown schematically in Fig.~\ref{fig:prism}. FFAs, although invented in the 1950's, have attracted considerable attention relatively recently due to their unique properties~\cite{Collot:2008zz}. In particular, they can provide lattices with very large acceptances in both the transverse and longitudinal planes, which is essential for muon beam applications. As muons are produced as tertiary beams, with large emittances and momentum spread, lattices for their transport and acceleration require large acceptance.

Our concept starts with a short muon bunch originating from the interaction of the compressed proton bunch with a target contained in a solenoid. The pions produced from proton interactions decay into muons, which are then injected into a small FFA ring. In this ring, equipped with the RF system, the longitudinal phase space rotation can be performed to transform the initial short muon bunch with a large momentum spread of $\sim\pm20\%$ into a long bunch with the momentum spread reduced by about an order-of-magnitude. The ring RF system can operate at harmonic number one, requiring low frequency RF cavities based on Magnetic Alloy technology, which allows to mix different RF frequencies to maximize the efficiency of the phase rotation~\cite{Ohmori:2008zza}. The narrow momentum spread beam is then extracted and sent to the experiment.

Several lattice solutions have been proposed for PRISM~\cite{Alekou:2013eta}, with an initial baseline based on a scaling DFD triplet~\cite{Sato:2008zze}, which was successfully constructed and verified experimentally~\cite{Witte:2012zza}. The studied solutions included both scaling and non-scaling FFA lattices, consisting of regular cells, and a racetrack geometry.

\begin{figure}
  \begin{center}
    \includegraphics[width=0.7\textwidth]{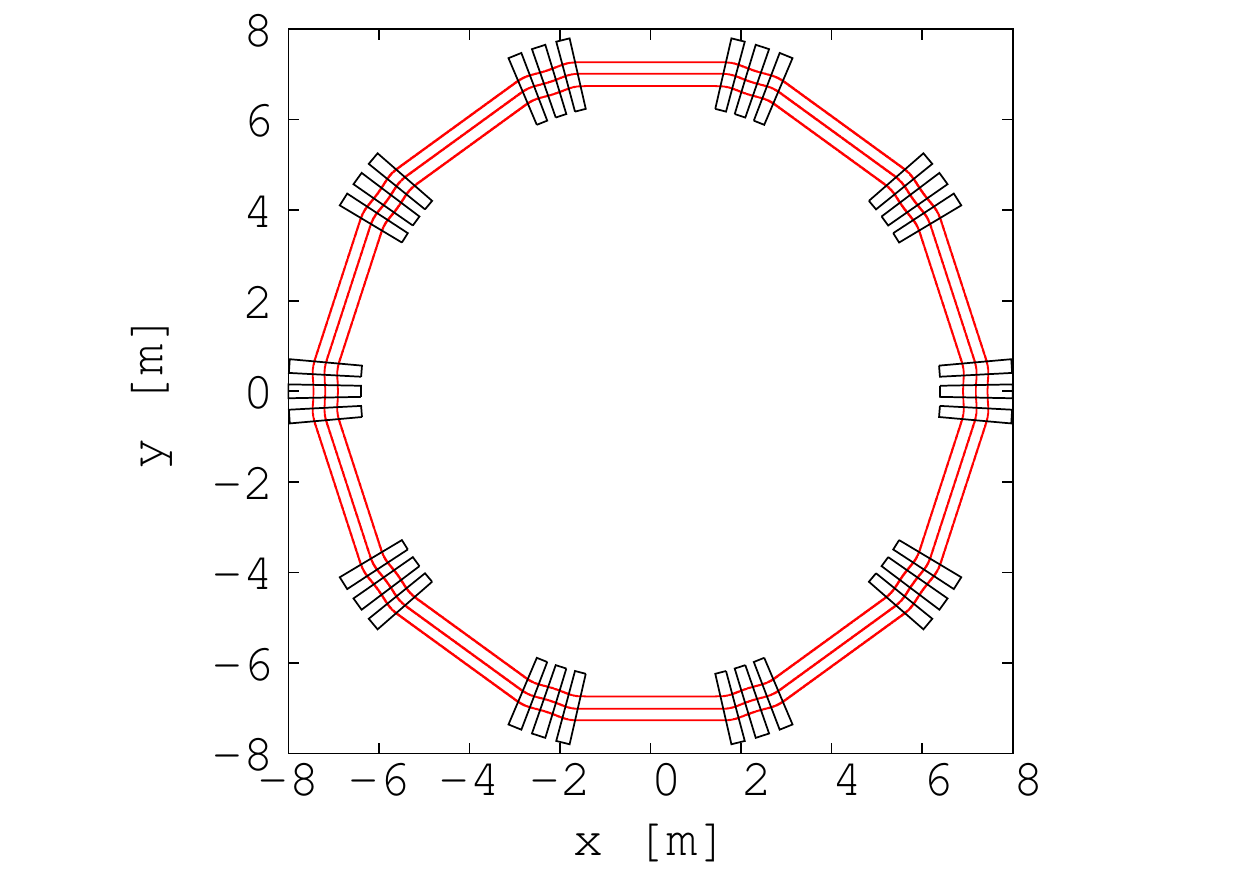}
  \end{center}
  \caption{
    Schematic drawing of the muon storage ring. The lattice is based on the scaling Fixed Field Alternating gradient (FFA) triplets with FDF symmetry. 
  }
  \label{Fig:FFA:ring}
\end{figure}

\begin{figure}
  \centering
    \includegraphics[width=0.45\textwidth]{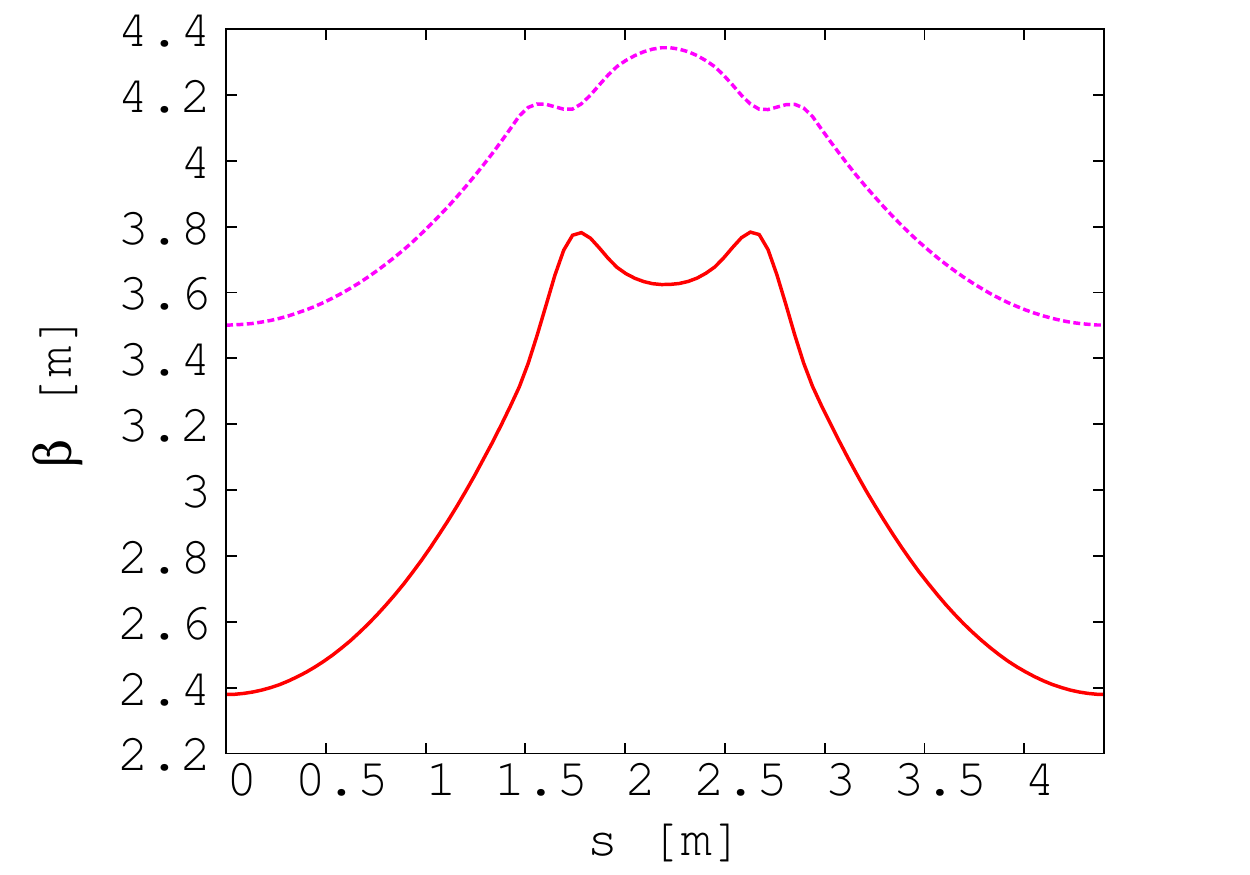}
       \includegraphics[width=0.45\textwidth]{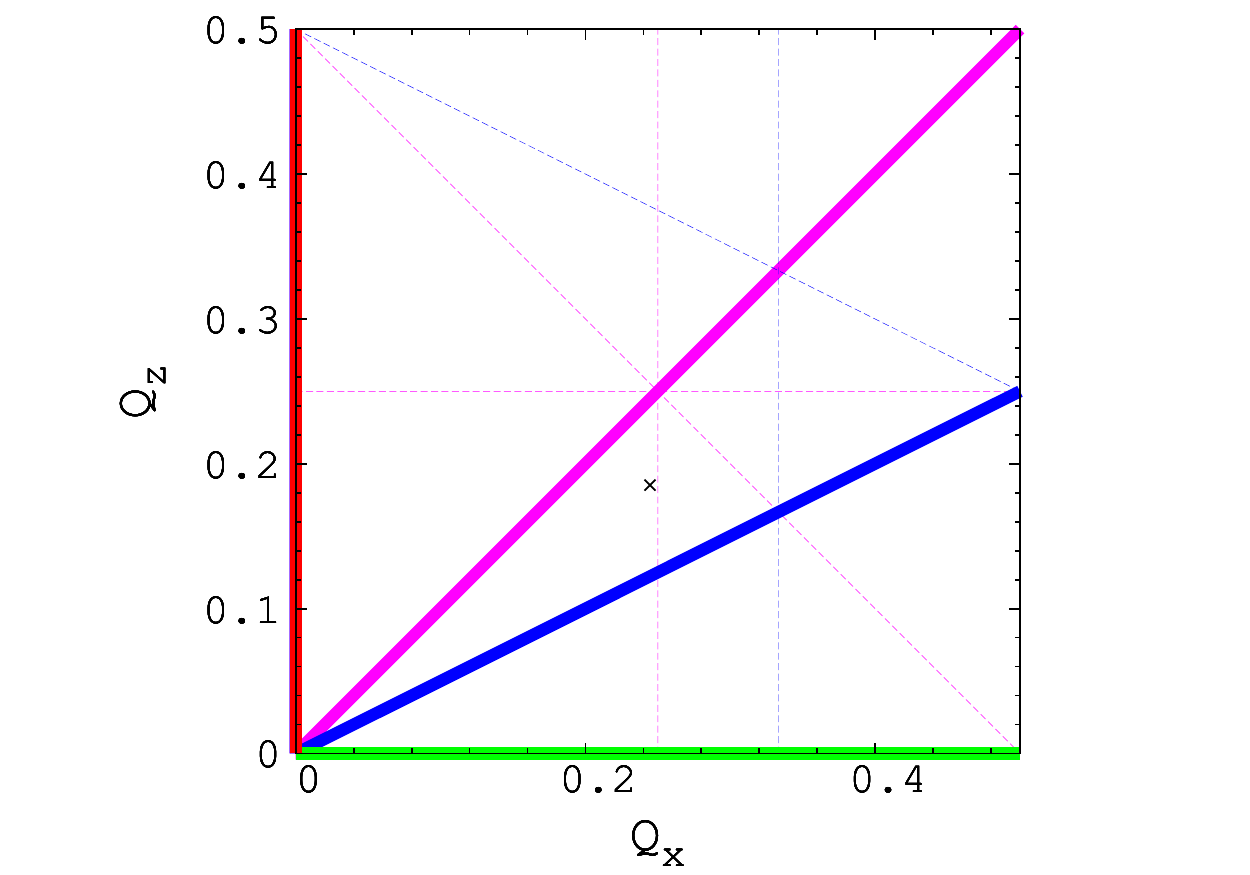}
\caption{Left: horizontal $\beta$ function (red curve) and the vertical one (purple curve) in a symmetric cell of the baseline FFA ring for PRISM. Right: working point (tune per cell) of the baseline FFA ring solution. The second and the third order resonance lines are represented by the blue and purple lines, respectively. The bold lines denote the systematic resonances.}
\label{Fig:FFA:optics}
\end{figure}

\begin{figure}
  \centering
    \includegraphics[width=0.45\textwidth]{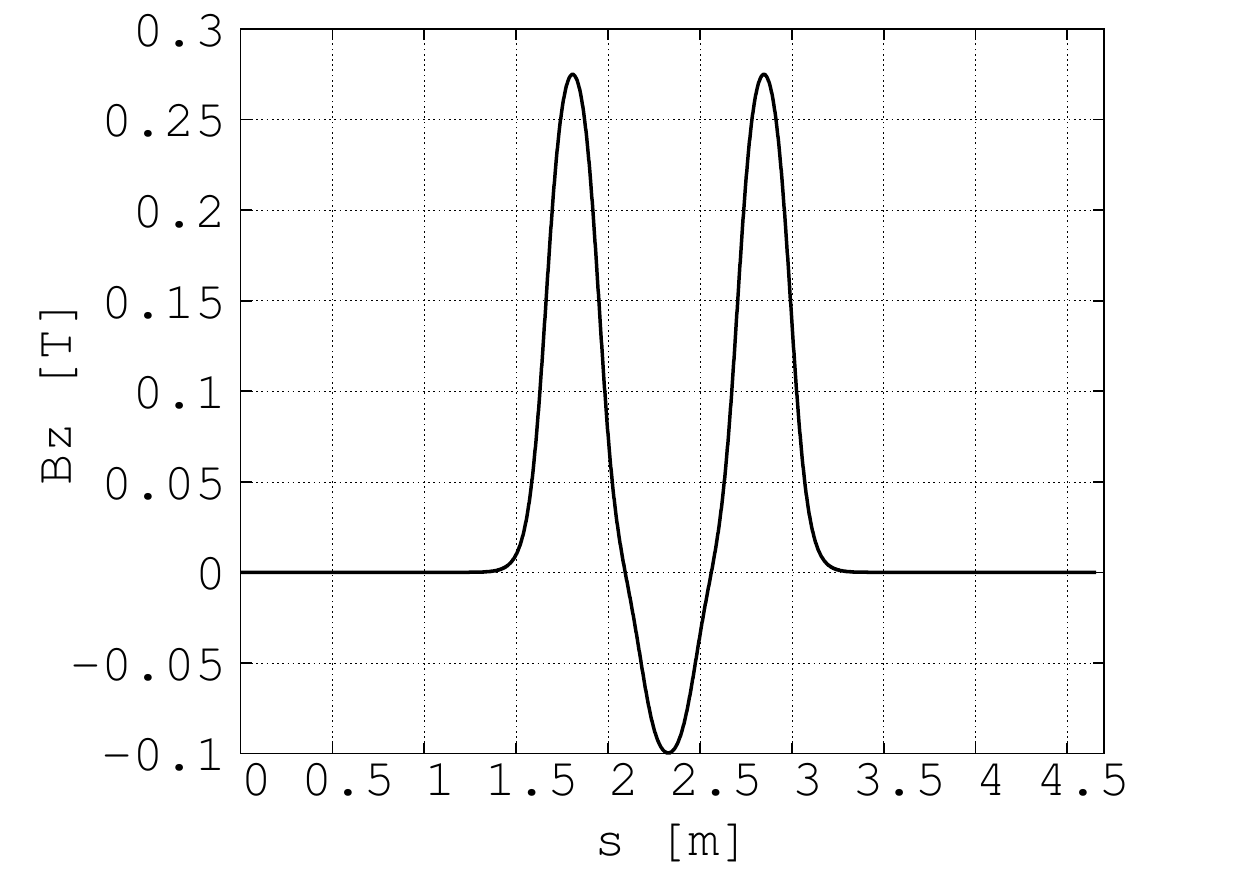}
       \includegraphics[width=0.45\textwidth]{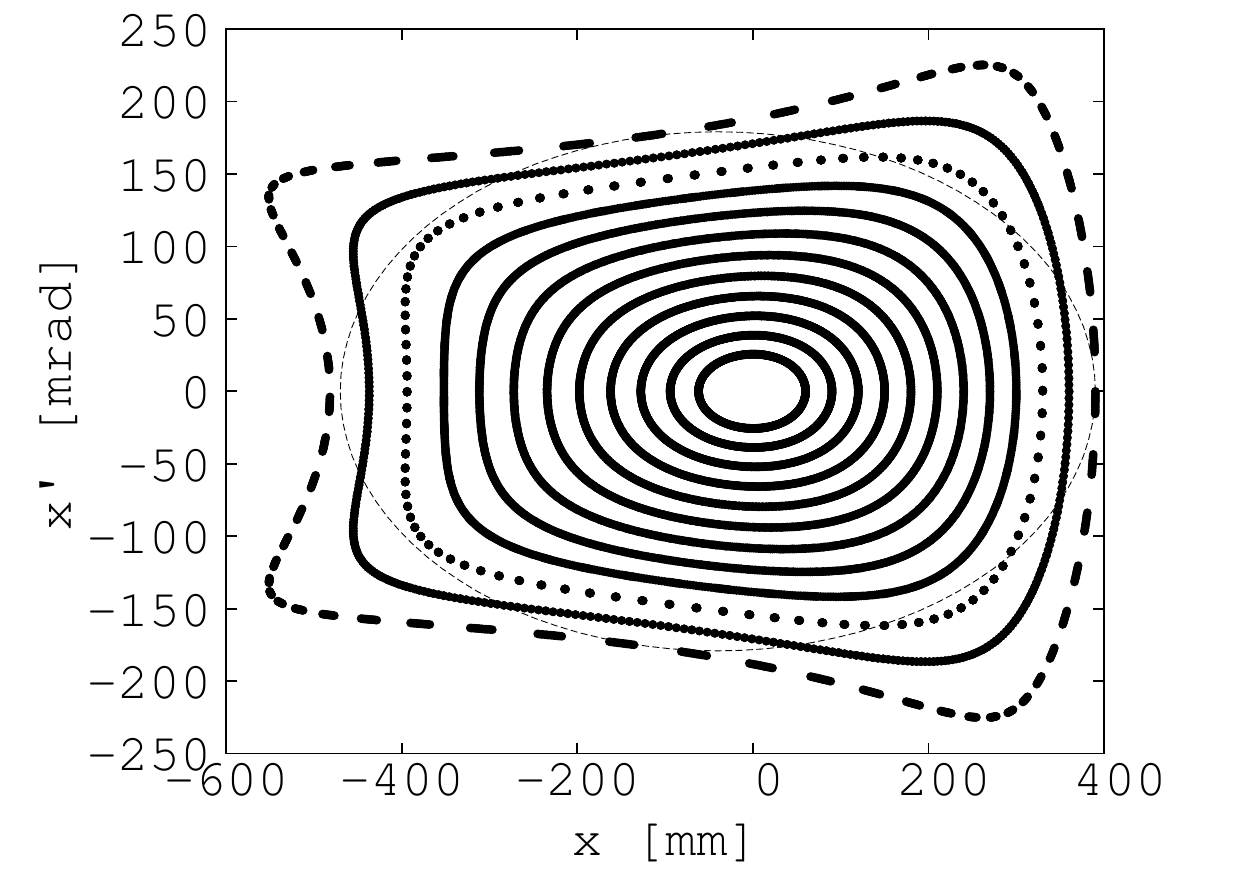}
\caption{Left: vertical magnetic field on the reference radius on the median plane of the baseline FFA ring. Right: horizontal dynamical acceptance studies in the FFA ring at the reference momentum. Particles are tracked over 100 turns with different amplitudes in the plane of study including a small off-set from the closed orbit in the other plane. The black ellipse represents the acceptance of $77\,\pi$\,mm\,rad.}
\label{Fig:FFA:tracking}
\end{figure}

The current baseline, proposed to take advantage of recent advances in FFA accelerators, is based on a regular FDF scaling lattice with ten identical cells, as shown in Fig.~\ref{Fig:FFA:ring}. The FDF symmetry can provide the required large dynamical acceptance, while simultaneously increasing the drift length for injection/extraction needs. The symmetrical optics in the ring with identical cells avoids large variations of the $\beta$ functions, which can drive dangerous resonances. The use of the scaling FFA principle makes possible --- thanks to its intrinsic zero-chromatic properties --- to keep the optics quasi-identical for off-momentum particles. In particular, the tune working point is  independent of momentum and situated away from the resonance lines, which can severely diminish the dynamical acceptance. The $\beta$ functions in one of the symmetric lattice cells and the working point (tune per cell) of the baseline FFA ring are shown in Fig.~\ref{Fig:FFA:optics}. The magnetic field on the median plane of the FFA ring along the reference radius --- described using the Enge model of the fringe fields --- is shown in Fig.~\ref{Fig:FFA:tracking}. The performance of the FFA ring was verified in tracking studies. In order to incorporate tracking through the combined-function magnets, taking into account the fringe fields and large amplitude effects, a code used for the full FFA machine developed previously (FixField code) was used~\cite{ Lagrange:2018triplet}. It is a step-wise tracking code based on Runge-Kutta integration, using Enge-type fringe fields. The results of the multiturn tracking shows that the horizontal dynamical acceptance of the machine is very large and exceeds an impressive figure of $77\,\pi$\,mm\, rad, as shown in Fig.~\ref{Fig:FFA:tracking}~(right-hand plot). The vertical dynamical acceptance is still being optimized, with the goal to achieve about $10\,\pi$\,mm\, rad. The vertical dynamical acceptance is typically smaller in horizontal FFAs, and the vertical physical acceptance is nevertheless limited by the injection needs. The main parameters of the baseline FFA ring solution for PRISM can be found in Table~\ref{tab:FFA:parameters}.

A significant challenge of the PRISM system resides in the design of the efficient beam transport from the decay solenoid, where the muon beam is formed, and its subsequent injection into the FFA ring. The beam in the solenoid is very strongly focused, has symmetric coupled optics in both transverse planes, and a large natural chromaticity. Dispersion is either zero or very small in either  a bending field or curved solenoids. This beam needs to be transported with negligible losses into the FFA, which has zero chromaticity, decoupled asymmetric optics with intermediate strength, and has relatively large dispersion function.

\begin{figure}
  \begin{center}
    \includegraphics[width=0.95\textwidth]{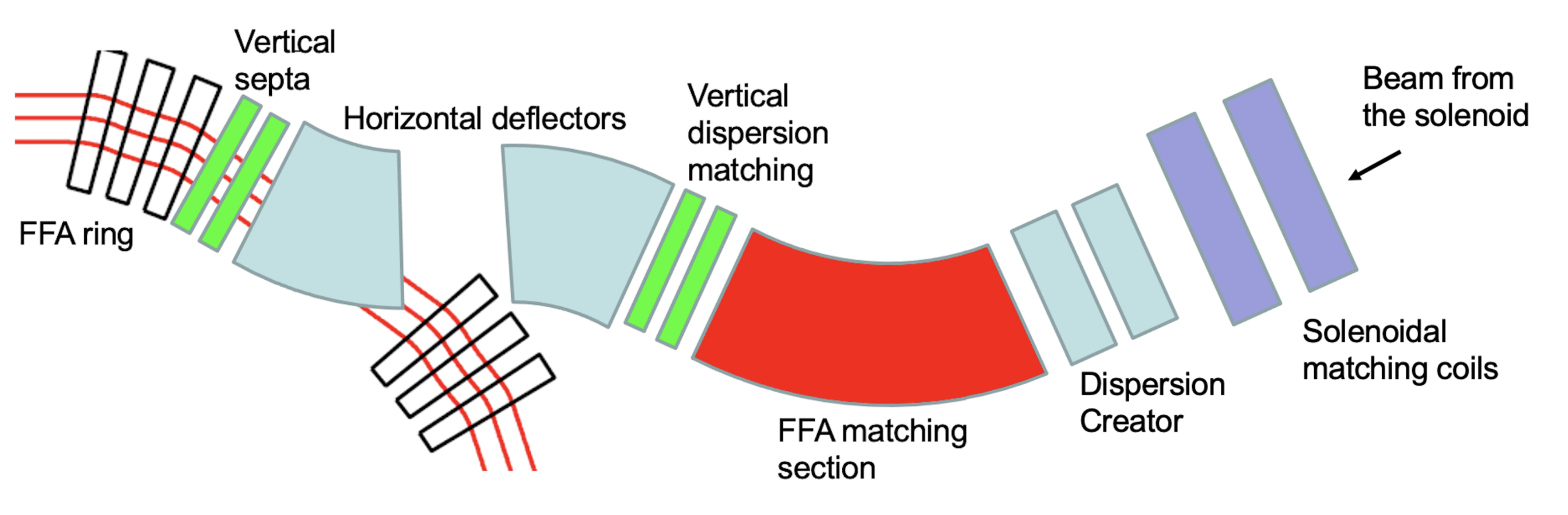}
  \end{center}
  \caption{
    Conceptual layout of the muon beam transport from the decay solenoid and injection to the FFA ring. The beam is moving from right to left. Injection kickers located in the ring are not shown.
  }
  \label{Fig:FFA:injection}
\end{figure}

A conceptual design for the beam transport and injection system has been proposed (Fig.~\ref{Fig:FFA:injection}). The beam first needs to exit the solenoid while the beam dynamics is under control. The proximity of the solenoid may also saturate the downstream Alternating Gradient (AG) iron dominated magnets, which should be avoided, and the reduction of the solenoid field needs to be controlled. A system of two solenoidal coils is proposed to perform the beam matching from the quasi-uniform decay solenoid into the downstream accelerator, controlling the beam size and divergence, while a more complex system may be required. Located downstream from this matching section, the dispersion creator consists of two rectangular dipoles with equal, but opposite bending angles, which generates the initial dispersion required for the FFA lattice. This is followed by the scaling FFA matching section, which aims to control the matching of the $\beta$ functions and the final dispersion into the values needed for the FFA ring. Finally, a system of bending magnets and septa is used to introduce the beam into the FFA ring with a vertical offset with respect to the circulating beam. The beam is bent vertically and the dispersion flips sign due to the strength of the bending magnets needed. The final horizontal magnet is a Lambertson-type septum~\cite{Paraliev:2018ski}. The horizontal septum may be followed by a vertical septum, which brings the beam into the ring. Additional vertical magnets upstream and downstream of the horizontal septum provides the matching of the vertical dispersion function to zero in the ring. The beam is then passed through one cell of the FFA, where the offset in position is transformed into a vertical divergence offset, which is cancelled by the kicker magnets finishing the injection process to place the beam on the circulating orbit. Further studies are needed to demonstrate the full feasibility of the described beam transport and injection system, and significant R\&D is required to address its full optimisation and to design the hardware. The extraction system and the beam transport from the FFA ring into the experiment can be realized as a reverse copy of the injection system, but may be significantly simpler as the momentum spread of the beam is a factor of ten smaller. The main challenge is the rise time of the extraction kicker(s), which needs to be shorter than the injection ones as the bunch at extraction is significantly longer.

The FFA can operate with both signs of muons, one at a time, if the machine fields can be reversed. However, it might be possible to apply the concept of the singlet FFA lattice~\cite{39289}, in which beams of both signs can circulate in the same direction simultaneously. Further studies are needed to investigate this idea.

\begin{table}
    \centering
\begin{tabular}{lcc}
\hline
\hline
Reference radius & 7~m\\
Length of one straight section & 3.15~m\\
Initial momentum spread & $\sim\pm$20\%\\
Final momentum spread & $\sim\pm$2\%\\
Reference muon momentum & 45~MeV/c (or lower)\\
Reference tunes per cell ($q_h$, $q_v$) & ($0.245$, $0.185$)\\
Number of cells in the ring & 10\\
Field index k & 4.3\\
Harmonic number & 1\\
\hline
\hline
\end{tabular} 
 \caption{Selected parameters of the FFA storage ring.}
 \label{tab:FFA:parameters}
\end{table}

%% file: Sections/linac.tex
\section{Induction linac}
\label{sec:linac}

Both the conversion and decay experiments need to stop the muon beams in as well-defined a volume as possible. The tradeoff is the fraction of stopping muons (and captured in the conversion experiments) vs.\ the stochastic $dE/dx$ loss and multiple scattering effects that worsen the resolution. The more material, the more muons stop from $dE/dx$, but as the amount of material increases the angular and energy resolutions correspondingly worsen. This becomes a fundamental limitation as the desired sensitivity of the experiment is increased. 

The PRISM task force settled on a reference momentum of 45 MeV/$c$ ($T = 9.2$ MeV)~\cite{Witte:2012zza}. This is a much larger momentum than that from a stopped muon beam (29.8 MeV/$c$, $T = 4.1\, \MeV$), and possibilities to lower that central momentum must be investigated, but an alternative would be to install an additional component to reduce the momentum of the beam that has been extracted from the FFA. For a number of reasons, traditional RF components are not well matched to this task, but an induction linac is a promising alternative.

Induction linacs are simple in principle~\cite{Mascureau1996INDUCTIONL}: an increase in current sets up an electric field through Faraday's Law, usually used to accelerare particles. In this case, however, the field will decelerate the muon beam.  An overview of induction linac technology~\cite{caporaso2000progress} shows that the needs of the new muon facility are within established capabilities, and they are in fact generally designed for much higher currents. However, additional R\&D would be required to arrive at a workable design.

%% file: Sections/MuDecay.tex
\section{Muon decay experiments}
\label{sec:mudecay}

The search for rare and forbidden muon decays takes advantage of intense muon beams stopped in a target to exploit the kinematical constraints of a decay at rest. These studies are only possible with positive muon beams, since negative muons would be captured by the target nuclei and distort the decay kinematics. The most recent experiments have exploited surface muons produced by the decay of pions at rest on the surface of a production target. The resulting $29.8~\MeV/c$ muons are further slowed down with a degrader to improve the stopping efficiency in a very thin target. However, multiple scattering deteriorates the momentum bite, partially negating the advantage of a lower beam energy. The stringent requirements on the reconstruction of low-energy electrons and photons pose severe, but somewhat different constraints on the experimental design for $\mu^+ \rightarrow e^+ \gamma$ and $\mu^+ \rightarrow e^+e^-e^+$ decays. We discuss each in turn.

\subsection{\texorpdfstring{$\mu^+ \to e^+ \gamma$}{mu to e gamma} decays}
The search for $\mu^+ \to e^+ \gamma$ in the decay of muons at rest is based on the reconstruction of a positron and a photon, emitted back-to-back from the target,each with an energy of $ 52.8~\MeV$. Given the two-body kinematics, the background rejection can independently exploit the positron and photon energies, their relative angle and time difference. At very high muon stopping rates, accidental backgrounds from two decays within the time resolution of the apparatus largely dominate over the intrinsic background from radiative $\mu^+ \to e^+ \overline \nu_\mu \nu_e \gamma$ decays. Since accidental backgrounds are proportional to the square of the beam intensity, $\Gamma_\mu^2$, as soon as the background yield $B$ becomes significant the overall sensitivity of the experiment starts scaling as $S/\sqrt{B} \propto \Gamma_\mu/\sqrt{\Gamma_\mu^2} = \mathrm{const}$. Hence, the resolutions and the efficiencies of the detectors, determining the expected background yield, set an optimal beam intensity, which can be beneficially increased only if the detector capabilities are improved.

\begin{figure}
    \begin{center}
    \includegraphics[width=0.45\textwidth]{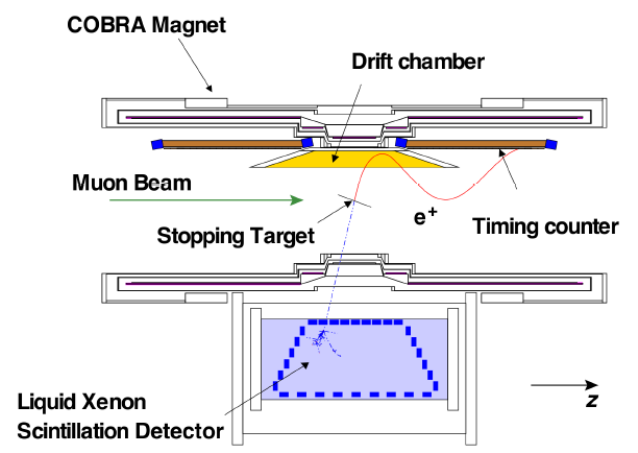}
    \includegraphics[width=0.45\textwidth]{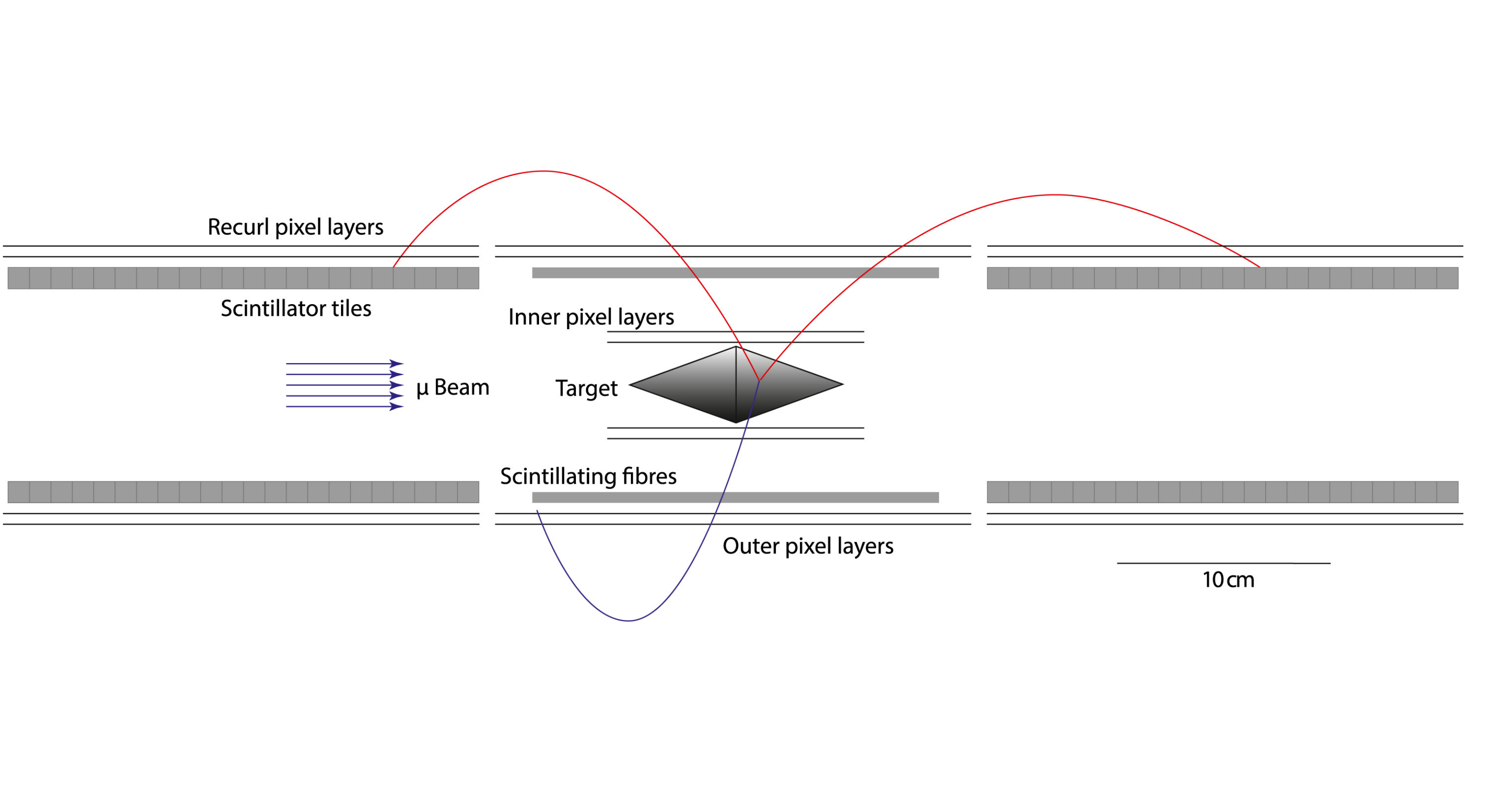}
    \end{center}
    \caption{A schematic of (left) the MEG experiment and (right) the Mu3e experiment. Adapted from Refs.~\cite{MEG:2016leq,Mu3e:2020gyw}.}
    \label{fig:megmu3e} 
\end{figure}

The best limit on this process has been set by the MEG experiment at PSI, $BR(\mu^+ \to e^+ \gamma) < 4.2 \times 10^{-13}$ at 90\% confidence level~\cite{MEG:2016leq}. The experiment, sketched in Fig.~\ref{fig:megmu3e}, reconstructed the positron kinematics with a set of 16 planar drift chamber in a graded solenoidal magnetic field. Gaseous detectors have the advantage of limiting multiple scattering in the detector material, and reducing the production of energetic photons from the annihilation of positrons in flight that could mimic the signal. The photon was reconstructed by a LXe detector instrumented with PMTs, measuring the photon energy, time and conversion point. The latter was combined with the intersection of the positron track with the planar target to determine the relative $e\gamma$ angle, under the assumption that the positron and photon come from the same production point. The MEG experiment has been recently upgraded with a new, single-volume, cylindrical drift chamber, the replacement of the PMTs in the inner face of the LXE detector with silicon photomultipliers and an increased granularity of the positron timing detector. The upgraded experiment is expected to take data for three years at $5 \times 10^7~\mu$/s, with a final sensitivity of $\sim 6 \times 10^{-14}$.

On-going studies show that that incremental improvements in photon calorimetry and positron tracking could push the limit well below $10^{-14}$, with an optimal beam rate of a few $10^8~\mu$/s~\cite{next_meg,Aiba:2021bxe}. Achieving this sensitivity requires, however, a  calorimeter with a much larger acceptance than the MEG LXe detector. At higher beam intensities, it could be advantageous to use thin layers of high-Z material to convert photons into $e^+e^-$ pairs, which could be accurately reconstructed to obtain a precise measurement of the photon energy and the position of the conversion point. Multiple conversion layers could be used to increase the total efficiency without worsening the resolution. The directional information provided by this technique, although insufficient to improve the $e\gamma$ angle resolution, would allow to reconstruct an approximate $e\gamma$ vertex to reject a significant fraction of accidental background events. The possibility to use multiple targets has also been considered, but they would have to be placed in a vacuum to avoid background from muon decays in the surrounding gas. 

In any scheme, positrons will need to be tracked in an extremely crowded environment with an extremely light detector. These conditions might be a challenge for gaseous detectors, which suffer from significant ageing effects --- the MEG-II drift chamber is expected to experience a gain loss of 25\% per year when operating at a muon stopping rate of $7 \times 10^7~\mu$/s~\cite{MEGII2018} --- and reduced reconstruction capabilities with high occupancy due to the limited granularity. A strategy to reduce the occupancy and pileup foresees an alternative arrangement of the wires in a drift chamber, using a transverse configuration similar to the Mu2e straw-tube detector instead of wires running parallel to the beam axis. This scheme would require a much larger number of wires, read out from the external surface of the detector, which cannot be too thick if positrons are to reach the timing detectors undisturbed. The possibility of collecting and detecting the photons emitted in the gas electron avalanches near the wires with optical fibres has been also proposed. This optical readout, intrinsically less affected by noise, could work with a reduced gas gain to slow down the ageing process. One could also consider replacing the wire chambers with time projection chambers (TPCs), but the large volume required to cover a wide angular range impose an unconventional geometry to avoid long drift distances. Contrary to typical devices, a TPC for $\mu^+ \to e^+ \gamma$ experiment would have a radial drift field, with electron multiplication and readout elements on the outer surface of the cylinder, exploiting the recent development of cylindrical micro-pattern gaseous detectors. Simulations show that the occupancy and the space charge inside the detector with a few $10^9~\mu$/s would be similar to what is expected in the GEM-TPC of the ALICE experiment, making it manageable although challenging. On the other hand, the necessity of keeping a low material budget poses severe constraints on the design of the readout. Alternatively, silicon pixel trackers could be a practical solution if the material budget can be kept under control. One potential candidate would be the $50~\mu$m HV-MAPS pixel sensor recently developed for the Mu3e experiment, which could become even more attractive if the thickness could be reduced to $25~\mu$m. While there is no shortage of ideas, an intensive R\&D program is needed to further assess the performance of these concepts and develop experimental proposals.

\begin{figure}
    \begin{center}
    \includegraphics[width=0.66\textwidth]{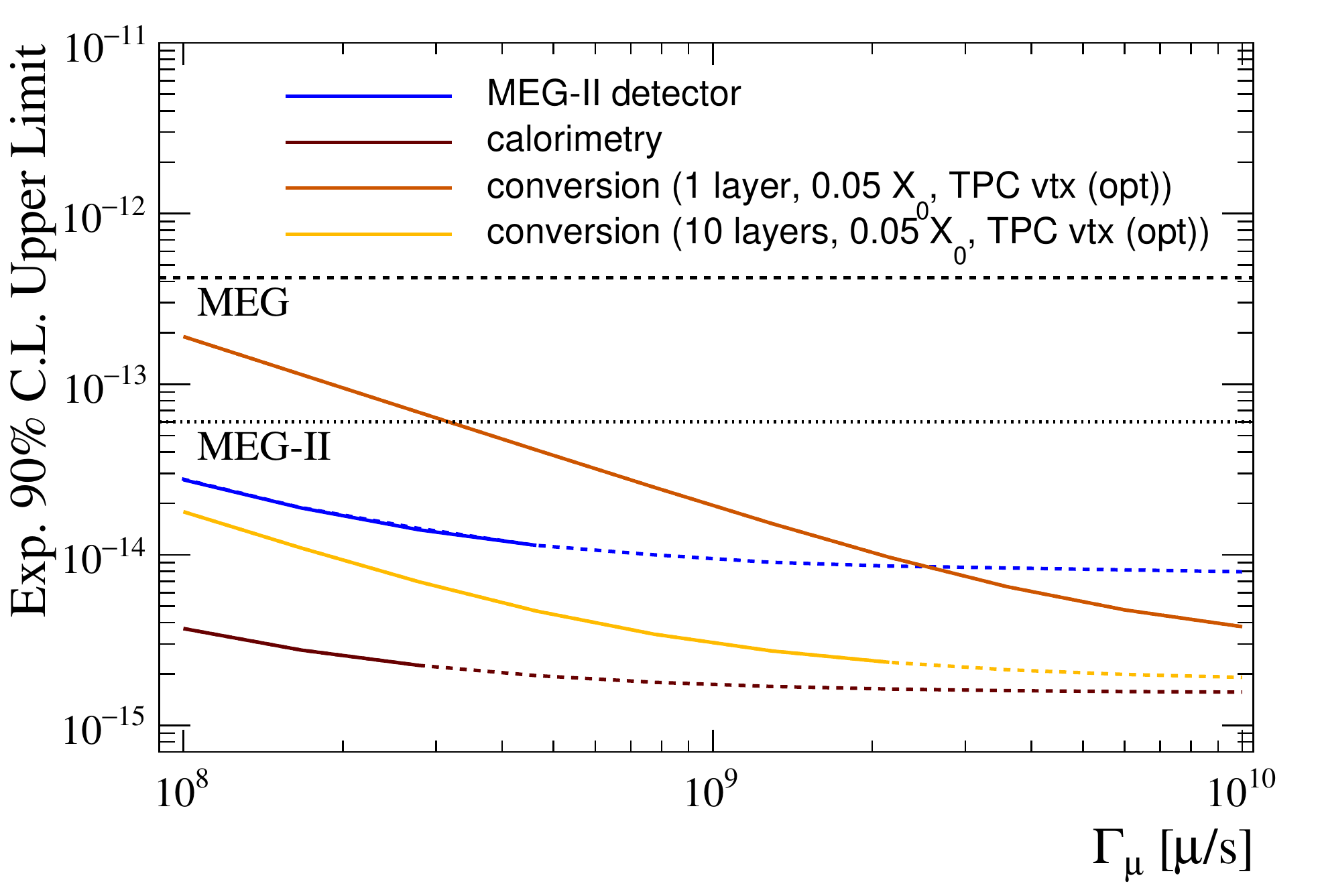}
    \end{center}
    \caption{Projected sensitivity for the next generation of $\mu^+ \to e^+ \gamma$ experiments as a function of the beam rate in different scenarios: with a detector having performances similar to the MEG-II design (blue); with photon calorimetry (dark brown); with photon conversion and a single (light brown) or a set of 10 conversion layers (yellow).}
    \label{fig:next_meg} 
\end{figure}

Predictions for a few different scenarios are shown in Fig.~\ref{fig:next_meg} as a function of the muon beam intensity. There are several avenues to reach a sensitivity down to $10^{-15}$: the calorimetric approach is generally better for lower beam intensities, while photon conversions take full advantage of a higher rate. Ultimately, the present techniques are still limited by uncertainties due to the positron interactions with the target and detector, and the energy resolution of the photon. Fully exploiting beam rates of $10^{10}~\mu$/s or more and breaking the $10^{-15}$ barrier will require a conceptually new experimental approach.

\subsection{\texorpdfstring{$\mu^+ \to e^+ e^+ e^-$}{mu to e+e+e-} decays}
The $\mu^+ \to e^+ e^+ e^-$ decay is identified by reconstructing three charged tracks originating from the same vertex. The accidental background, coming from muons decaying in different positions, is consequently strongly suppressed, and the dominant background arise from the SM $\mu^+ \to e^+ e^+ e^- \overline \nu_\mu \nu_e$ decay up to extremely high beam rates. Consequently, the beam rate can be increased beneficially, almost irrespective of the achieved resolutions up to a certain point.

The current limit has been set by the SINDRUM experiment in 1988, BR$(\mu^+ \to e^+ e^+ e^-) < 1.0 \times 10^{-12}$ at 90\% CL~\cite{SINDRUM:1987nra}. The Mu3e experiment at PSI~\cite{Mu3e:2020gyw}, currently under commissioning, plan to improve the sensitivity by several orders of magnitude. The experimental apparatus, depicted in Fig.~\ref{fig:megmu3e}, includes four layers of HV-MAPS pixel detectors and scintillating fibres and tiles for a precise timing of the three particles. A Phase-I experiment will be performed on the same beam line as the MEG experiment, with a maximum beam rate of $\sim 10^8~\mu$/s, to achieve a sensitivity of $\sim 10^{-15}$. A Phase-II detector with additional tracking and timing stations could take full advantage of the $10^{10}~\mu$/s foreseen at the proposed HIMB facility at PSI to further improve the sensitivity by an order-of-magnitude. 

A critical aspect of the design and operation of $\mu^+ \to e^+ e^+ e^-$ experiments at high intensities is the achievable beam-spot size. In order to keep the accidental background level acceptable at $10^{10}~\mu$/s, the first silicon layer should be as close as possible to the decay vertices, since the background rejection power scales with the inverse of the beam diameter squared. The emittance of the HIMB line is expected to be one order of magnitude larger than that of the PSI beam line used in phase-I. Several options could be envisioned to decrease the beam spot size at the center of the detector solenoid: reduce the muon momentum, increase the magnetic field or design a dedicated optics. Any new high-intensity muon beamline designed for this type of searches should take these aspects into consideration.

Interestingly, a $\mu^+ \to e^+ e^+ e^-$ experiment could also search for $\mu^+ \to e^+ \gamma$ decays with the inclusion of one conversion layer in a suitable position. Despite the relatively poor efficiency offered with a single layer and a lack of optimization for the two-body kinematics, this type of experiment could still improve upon MEG-II with a very high beam rate. More remarkably, it would represent an important step toward the subsequent generation of $\mu^+ \to e^+ \gamma$ searches.

\subsection{Other channels}
The last few years have seen a renewed interest in low-mass particles as potential dark matter candidates or mediators between the SM and the dark sector. Neutral particles lighter than the muon could be searched for in muon decays, either as missing energy or through their decay into SM particles. 

In the case of missing energy, the golden channels are $\mu^+ \to e^+ X$ and $\mu^+ \to e^+ \gamma X$, where the new particle $X$ escapes detection but emerges as a peak in the distribution of the missing reconstructed mass. The steep edge of the $\mu^+ \to e^+ \overline \nu_\mu \nu_e$ energy spectrum at the kinematic end point makes these searches extremely challenging when the invisible particle is close to massless, since a mis-calibration of the positron energy scale can easily mimic the signal. On the other hand, since the $V-A$ structure of the weak interaction suppresses the spectrum in the direction opposite to the muon spin, polarized muons could potentially boost the sensitivity if the NP decays are controlled by a different structure
(e.g.\ $V+A$)~\cite{megII-fwd}. The latest limit on $\mu^+ \to e^+ X$ decays come from the TWIST experiment~\cite{twist_X}, but the best bounds under the $V+A$ hypothesis dates back to 1988~\cite{jodidio} and exploited the aforementioned suppression of the SM muon decay spectrum. Dedicated experiments at very high muon beam rates  exploiting the most recent technologies can  significant improvements of the sensitivity. 

Inclusive searches can also be performed for new particle decays into visible states, such as $\mu^+ \to e^+ X$ with either $X \to \gamma \gamma$ or $X \to e^+ e^-$. While the first channel was recently searched for by MEG~\cite{meg2g}, the second one is under study at Mu3e, and dedicated experiments could be envisioned at future facilities as well. Similar considerations hold for the search of $\mu^+ \to e^+ \gamma \gamma$ with no intermediate state. The acceptance of MEG and MEG-II is strongly suppressed by the geometry and the trigger, which are optimized for their two-body kinematics. Nonetheless, technologies similar to those adopted in $\mu^+ \to e^+ \gamma$ experiments could be deployed in a dedicated experiment, which would provide increased sensitivity to couplings poorly probed by other muon CLFV searches.

%% file: Sections/MuConversion.tex
\section{Muon conversion experiments}
\label{sec:conversion}

The search for $\mu^-N - e^-N$ conversion is carried out by stopping a negative muon beam in a target, wherein muons are captured in atomic orbits and converted into electrons through a coherent interaction with the nucleus. Since the nucleus remains unchanged during this process, the energy of the outgoing electron (CE) is close to the muon rest mass:
$$ E_{CE} = m_\mu{}c^2 - E_b - E_\text{recoil} $$
where $E_b$ is the binding energy of the muon in the $1S$ orbit and $E_\text{recoil}$ is the energy of the recoiling nucleus. For Al, the conversion energy is $E_{CE}=104.9~ \MeV$.

\begin{figure}[hb]
    \begin{center}
    \includegraphics[width=0.45\textwidth]{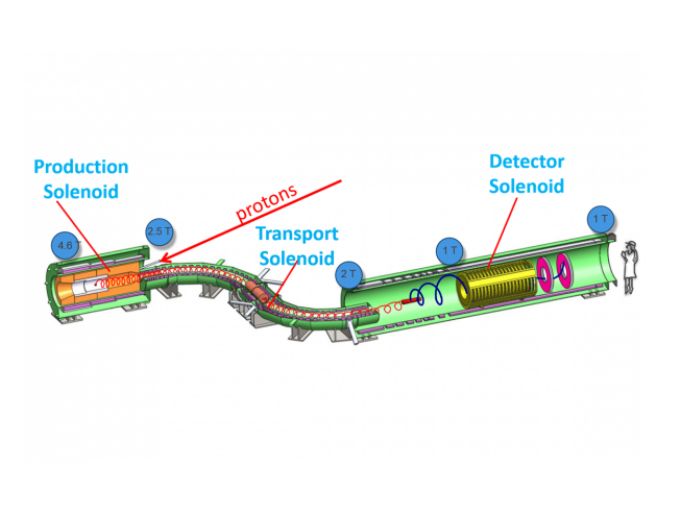}
    \includegraphics[width=0.45\textwidth]{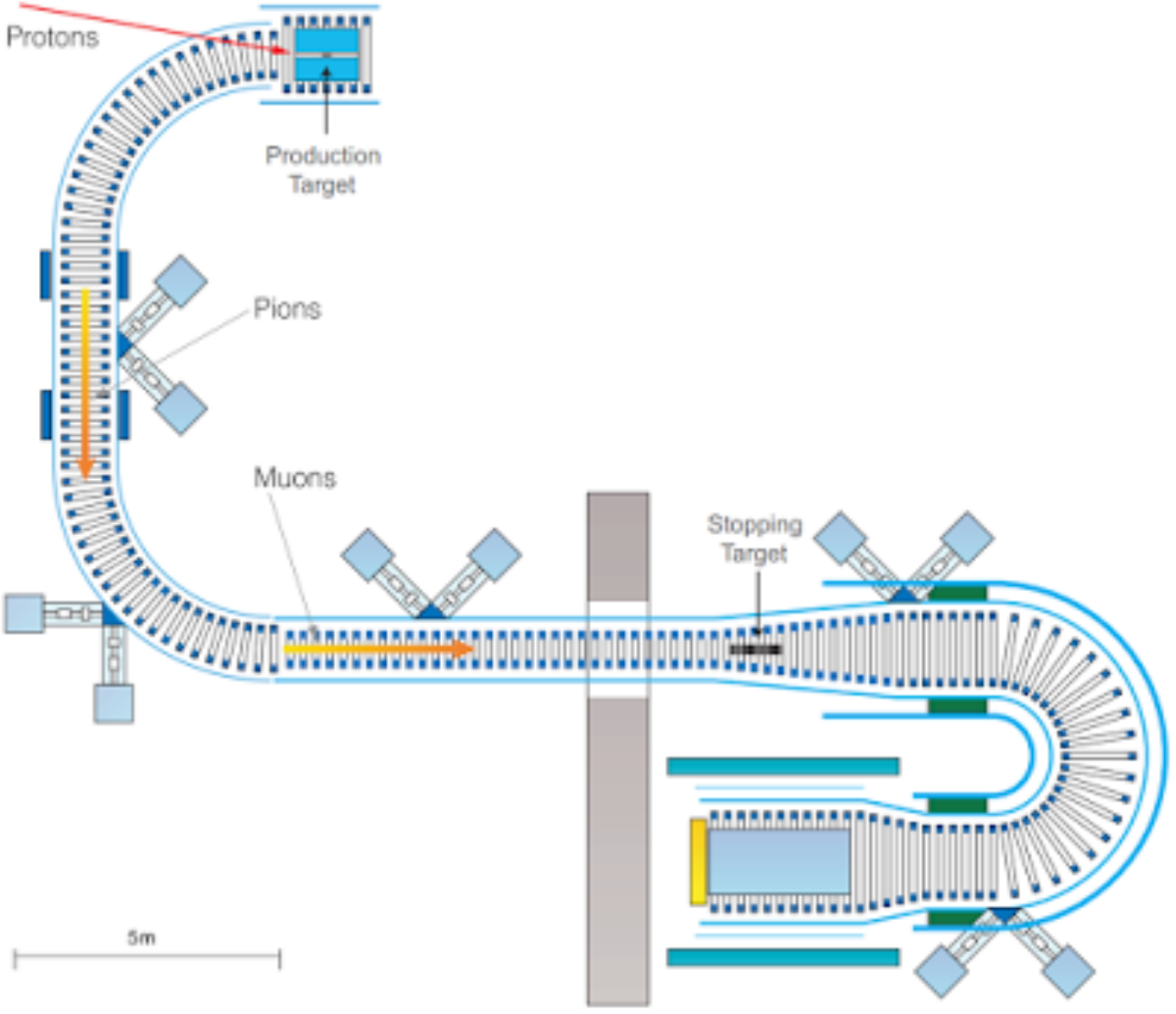}
    \end{center}
    \caption{A schematic of (left) the Mu2e experiment and (right) the COMET experiment. Adapted from Refs.~\cite{comet-tdr,bartoszek2015mu2e}.}
    \label{fig:mu2ecomet} 
\end{figure}

The best limit on this process has been set by the SINDRUM II experiment at PSI on a gold target, $R_{\mu e}(Au) < 7 \times 10^{-13}$ at 90\% CL~\cite{SINDRUMII:2006dvw}, where $R_{\mu e}$ is the muon conversion rate normalized to the muon nuclear capture rate. The next generation of experiment, Mu2e at FNAL~\cite{bartoszek2015mu2e} and COMET at J-PARC~\cite{comet-tdr}, are expected to improve this sensitivity by about four orders of magnitude. Both experiments will use pulsed proton beams to form an intense muon beam, transported onto a stopping target. The conversion electron will be identified with a high-resolution tracking system and a calorimeter placed in a solenoid. The full experimental layouts are illustrated in Fig.~\ref{fig:mu2ecomet}. The transport line is one of the most characteristic features of these setups, based on curved solenoids to shield the detector from the direct line of sight of the production target and select negatively charged muons (the helical trajectories of charged particles are deflected in opposite directions in the vertical plane, and the resulting drift can be used to filter wrong-sign particles). In the Mu2e configuration, the stopping target sits directly in front of the detector system. This design has the advantage of being charge symmetric, enabling the search for $\mu^-N  \rightarrow e^+ N'$ decays and measure positrons from RPCs, but the innermost regions of the detector must be left uninstrumented to handle the large current generated by low-momentum particles. By contrast, COMET uses another bent solenoid to downstream of the stopping target before a tracking detector to limit the acceptance to electrons with momenta near that of the expected signal. This allows for a fully instrumented volume, but at the cost of charge symmetry.

The energy and trajectory of the conversion electron must be measured with the best possible resolution to reject intrinsic background due to electrons produced from muon decay-in-orbit (DIO). The latter have a very high rate, and a kinematic endpoint equivalent to the CE energy. Beam-induced backgrounds, such as beam electrons and radiative pion capture (RPC) in the muon stopping target, must also be suppressed. The current generation of experiments take advantage of the pulsed proton beam structure and the relatively short RPC lifetime to delay the search for the conversion signal until the backgrounds reach negligible levels. This approach imposes stringent constraints on intra-pulse beam extinction,  a factor $10^{-10}$ in Mu2e or COMET, and limits the choice of target materials with long atomic muon lifetimes. As previously shown in Fig.~\ref{fig:sens-and-lifetime-vs-z}, the sensitivity to various BSM scenarios and the mean lifetime of a muonic atom depend on the target atomic number $Z$ in opposing ways. The COMET and Mu2e experiments, which have similar proton beam structures with pulse lengths of 100--200 ns and pulse frequencies of 1.1--1.7 $\mu$s, use a stopping target made of aluminum with a muonic atom lifetime of 864 ns. Heavier targets, which would be needed to study a potential signal, are out of reach. Finally, cosmic rays can mimic signal events in several ways, including the production of an electron with the conversion energy originating in the stopping target, and are suppressed with a dedicated veto system. 

The capabilities of AMF present a significant opportunity for a next-generation experiment to search for $\mu^-N - e^-N$ conversion. The long muon storage time will drastically reduce the impact of beam-induced backgrounds, enabling the use of high-$Z$ target material. The increase of the primary proton beam power, in conjunction with the muon cooling capabilities of the FFA, will also greatly increase the number of stopped muons in the target, ideally providing the opportunity to probe rates down to $10^{-19}$ or lower. 

\begin{figure}[ht]
    \centering
    \includegraphics[width=0.6\textwidth]{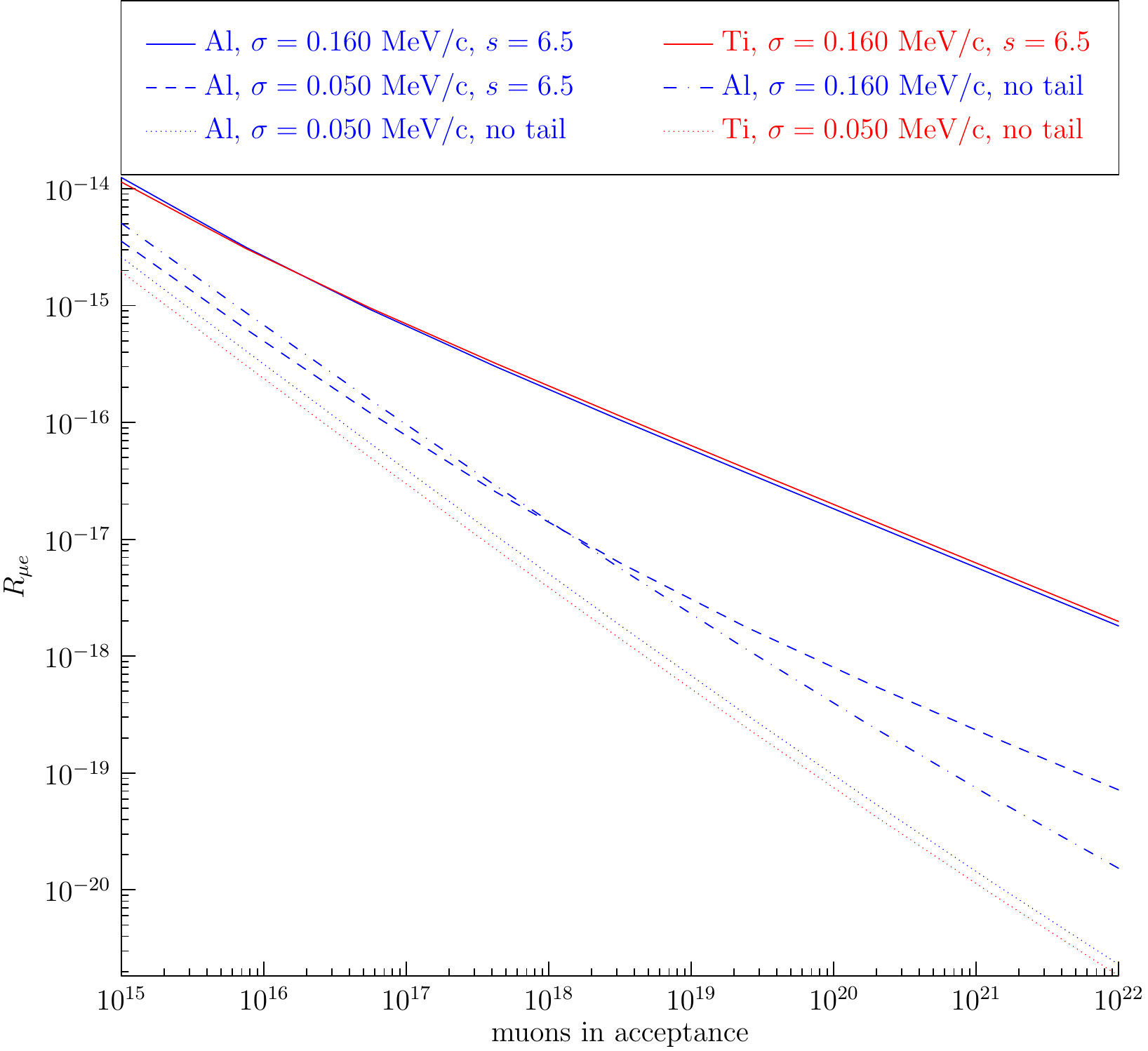}
    \caption{Median $5\sigma$ discovery sensitivity scaling with
        stopped muon statistics for different experimental resolutions: ``Mu2e-like'' solid lines with core resolution of $0.160$~MeV/c (Landau FWHM of $0.377$~MeV/c) and high side power tail $(p-p_{tail})^{-s}$ with $s=6.5$, improved core resolution or eliminated power tail, and both improved core resolution and eliminated power tail \cite{sensitivity-scaling}}
    \label{fig:sensitivity-scaling}
\end{figure}

This level of sensitivity requires an exquisite momentum resolution to achieve an optimal background rejection, especially for the DIO contribution, which scales with the number of stopped muons. Under the assumption that all other background sources are negligible, the achievable discovery sensitivity is illustrated in Fig.~\ref{fig:sensitivity-scaling} for different resolution functions. Improvements to the``Mu2e-like'' resolution  are clearly needed to make progress. This could be partly achieved by reducing the amount of target material, since energy loss fluctuations inside the target contribute to a substantial fraction of the resolution. The muon beam at the exit of the FFA has a lower momentum distribution than the muons reaching the Mu2e stopping target, and would therefore require less material to be stopped. Cosmic rays are also a significant source of background events, but recent studies of a proposed Mu2e-II experiment have shown that this level of performance is within reach~\cite{mu2eii:WP}.

\begin{figure}[ht]
    \includegraphics[width=0.6\textwidth]{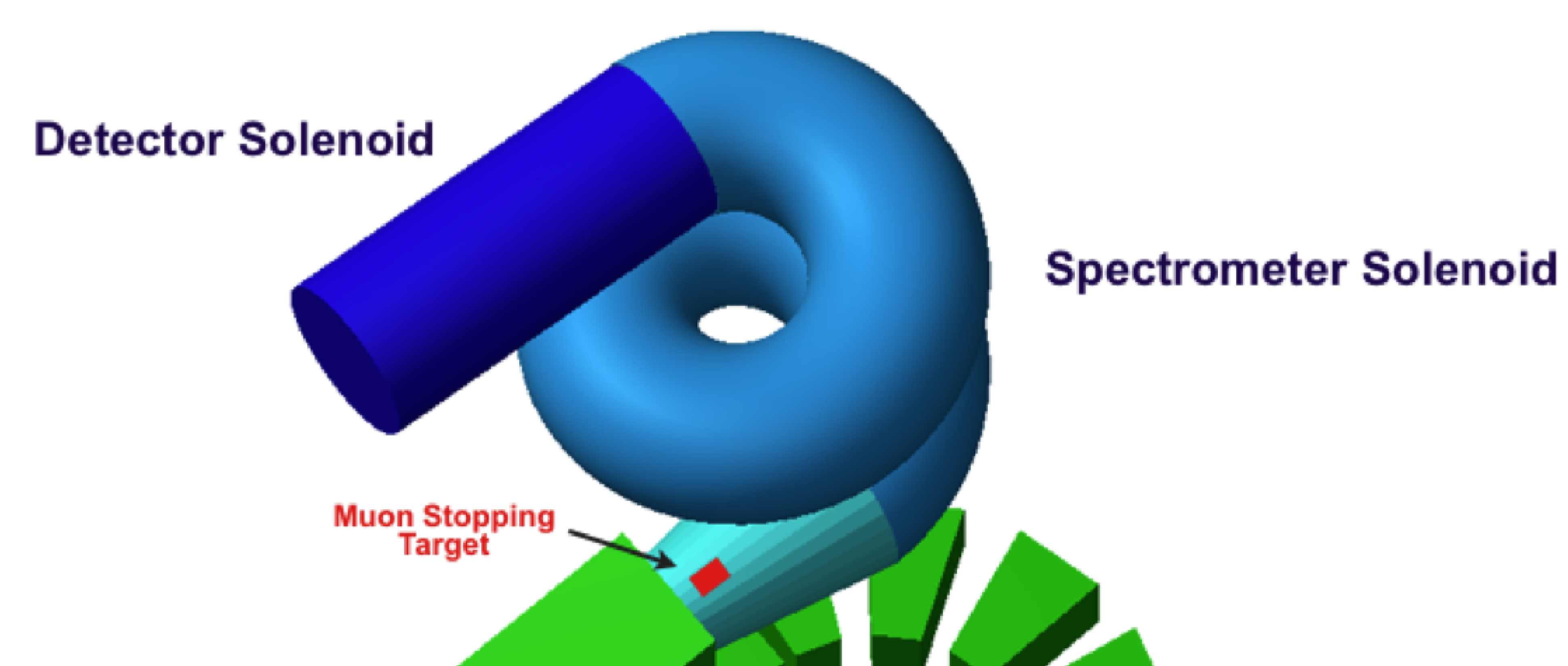}
    \caption{PRIME/PRISM electron spectrometer scheme, adapted from Ref.~\cite{KUNO2005376}.}
    \label{fig:prism-prime_spectrometer}
\end{figure}

Addressing these challenges could be accomplished by extending the COMET approach to a larger spectrometer to reduce the sensitivity to low-momentum backgrounds, as proposed by the PRISM/PRIME experiment~\cite{KUNO2005376}. A recent update suggests that a Spectrometer Solenoid with curvature beyond $360\degree$ may be required in order to probe conversion rates below $10^{-18}$~\cite{SATO:2020}. A Spectrometer Solenoid scheme featuring $540\degree$ of curvature, shown in Fig.~\ref{fig:prism-prime_spectrometer}, would only select electrons with momentum near the CE signal and provide the required background rejection. The lack of charge symmetry in this design isn't critical since beam induced backgrounds are already suppressed by the FFA. 

A number of promising candidates for high-performance, low-mass trackers exist, including the proposed Mu2e-II straw-tracker with 8 $\mu$m wall thickness~\cite{mu2eii:WP}, low-mass silicon sensors, such as HVMaps or micro-pattern gas detectors proposed for the Belle-II tracking TPC~\cite{belle2:WP}, and even exotic novel materials~\cite{novel-sensors:WP}. Potential technologies are also available for the calorimeter and for the cosmic-ray veto, both demonstrated by the Mu2e-II effort~\cite{mu2eii:WP}. While a significant R\&D effort is needed, the development of a detector concept taking full advantage of AMF should be a manageable  challenge. 

%% file: Sections/synergies.tex
\section{Synergies}
\label{sec:synergies}
This section briefly reviews synergies with a potential dark matter program at FNAL, as well as efforts in high power targeting in a solenoid.

\subsection{Dark Matter Program}
The compressor ring could also be used to rebunch the PIP-II beam for an accelerator-based dark matter experiment~\cite{boosterDM}. This experiment needs a higher-intensity, lower repetition rate beam than that envisioned for AMF. Potential operating modes, under the assumption of a 100 m circumference $0.8~\GeV$ ring, are detailed in Table~\ref{tab:compressorOptions}. In the case of AMF, the ring is filled with 16 bunches evenly spaced, separated by 24 ns. A kicker fires every 100 Hz (may be the same kicker device or they might alternate firing), removing exactly one of the bunches at a time. The resulting pulse structure is $8 \times 10^{13}$ protons over 12 ns every 100 Hz. The mode of operation for dark matter searches follows a similar pattern.

The construction of a suitable compressor ring would position Fermilab to build a world-class physics program in two significant efforts in the Rare and Precision Frontier: charged lepton flavor violation and dark matter at accelerators.  

\begin{table}[ht]
\begin{center}
\begin{tabular}{|l|r|r|r|}\hline
Description&Protons-Per-Pulse&Pulse Spacing (ns) & Repetition Rate (Hz)\\\hline
AMF              & $7.8 \times 10^{13}$ & 24  &   100\\
Dark Matter      & $6.2 \times 10^{14}$ & 196 &   100\\\hline
\end{tabular}
\caption{Potential operating modes for the compressor ring}
\label{tab:compressorOptions}
\end{center}
\end{table}

\subsection{High Power Targeting in a Solenoid}

Muon collider proponents have been studying how to achieve $\mathcal{O} (1~\rm MW)$ on a target inside a solenoid since at least the 1990's. The joint AF/NF/RP targetry workshop summarized the muon collider problems: 1--4 MW at 5--10 GeV. The muon collider originally considered mercury targets but switched to carbon~\cite{targetWork,calvianiTalk} since mercury presents significant challenges~\cite{pellemoineTalk}, including:
\begin{itemize}
    \item target lifetime. The SNS target requires 2--3 changes per year.
    \item environmental, the Minimata Convention would make it difficult to irradiate large amounts of mercury~\cite{minimata}.
\end{itemize}
    
Some of the challenges associated with such a target have been outlined in Section~\ref{sec:target}. Every one of them will occur in some form in the muon collider, and the construction of such a system for AMF would instruct the design and construction of such a collider.

%% file: Sections/Conclusion.tex
\section{Conclusion}
\label{sec:conclusion}
This report outlines the potential of the Advanced Muon Facility at Fermilab, a new complex to provide the world's most intense positive  and negative muon beams. This facility would enable a broad muon physics program, in particular next generation searches for charged lepton flavor violation. The physics case of CLFV is extremely compelling, and an observation would be a clear sign of NP, opening a window on the mechanism generating neutrino masses, and more generally GUT-scale dynamics. Constraints on muon CLFV are already impressive: $\mu \rightarrow e \gamma < 10^{-14}$ and $\mu \rightarrow 3e$ will be investigated at similar levels, and muon-to-electron conversion will be studied at Mu2e and COMET to ${\cal  O} (10^{-17})$. This facility could push two orders of magnitude past these limits, down to $10^{-19}$ for muon conversion, or study potential signals in great detail. Such sensitivity will probe NP mass scales up to a few  $10^4~\TeV$, setting exceptionally stringent constraints on physics beyond the SM and/or present new ``fine-tuning" problems for models. There are also clear synergies with the development of a muon collider, and a Dark Matter program (with its own strong physics case). A single facility and suite of experiments will create a community of several hundreds of physicists exploiting the PIP-II facility at Fermilab. 

This document lays out the first steps towards the realization of this facility, but it will take a small team of accelerator scientists working in conjunction with experiment designers to develop those ideas. No insurmountable issues have been identified so far. The objective of this white paper is to obtain P5 endorsement to pursue the studies necessary to produce a detailed proposal at Fermilab, ideally a CD-1 level document that would be studied at the next P5. The cost of such commitment is modest, but the potential payoff is enormous. The most expensive piece of equipment, the PIP-II complex, is already underway, but only 1\% of its power will be used by the future neutrino program. This proposal represents a unique opportunity to fully exploit the remaining capabilities with a world class physics program. This program has no competition, providing unique, compelling new physics searches that cannot be performed in any other way.